\documentclass{aastex}
\usepackage{spr-astr-addons}
\usepackage{url}
\usepackage{natbib}
\usepackage{soul}
\usepackage{pdflscape}
\usepackage{adjustbox}
\usepackage{array} 
\RequirePackage{color}
\begin{document}

\title{Asymmetric Cross-Correlation functions with delays in Sco X-1: Evidence of possible Jet triggering}
\slugcomment{Not to appear in Nonlearned J., 45.}
\shorttitle{Lags in Sco x-1}
\shortauthors{Gouse, Abhishek, and Sriram}

\author{SD Gouse, \altaffilmark{1}} \email{syedgouse@osmania.ac.in}
\author{ M. V. R. Abhishek,  \altaffilmark{1}} \author{K. Sriram \altaffilmark{1}}
\affil{syedgouse@osmania.ac.in, Department of Astronomy, Osmania University, Hyderabad, 500007, India}

\email{syedgouse@osmania.ac.in}


\begin{abstract}
The formation and origin of jets in Z sources are not understood very well, although a strong X-ray-radio correlation has been noticed. We analyzed seventeen observations of Sco X-1 observed by the Rossi X-ray Timing Explorer (RXTE) where radio jet emissions were detected. In five observations, we report the detection of asymmetric cross-correlation functions (CCF) with delays of a few tens of seconds between soft and hard energy bands light curves in the horizontal branch associated with a flat-topped noise in the power density spectrum (PDS). Interestingly, these five observations were connected to a ballistic-type radio emission. We performed simulations to confirm and robust the cross-correlation coefficients and the observed lags. The CCF was highly symmetric in the remaining twelve observations, exhibiting NBO (Normal Branch Oscillations) or NBO+HBO (Horizontal Branch Oscillations) in the PDS. During these X-ray observations, the radio observations were found to be associated with an ultra-relativistic flow (URF) radio emission. The X-ray spectrum analysis of the two observations that showed core radio emission and abrupt variations in the PDS and CCF revealed that the bbody fractional flux varied by 10–20\%, but the spectral parameters did not vary.

 We suggest that the ballistic jet might have triggered the instability in the inner region of the accretion disk, viz., the boundary layer (BL) plausibly along with the corona, causing the asymmetry with delays observed in the CCFs, and it also explains the absence of any oscillation features in the PDS, leaving behind a flat-topped noise. During symmetric CCF, the accretion flow was steady, hence, NBO / NBO+HBO was persistent. However, connecting the URF either to NBO or HBO is difficult since the majority of PDS exhibit an NBO alone rather than a NBO+HBO. We hypothesize that URF is most likely related to the phenomenon that causes NBO. Overall, we conclude that the asymmetric CCF shows that the inner part of the accretion disk is unstable due to the triggering of a ballistic jet and constrains the inner accretion region's size to about 20-30 km, which possibly causes the accretion ejection.

\end{abstract}


\section{Introduction}

Low Mass X-ray Binaries (LMXBs) are a class of X-ray binaries that have a weakly magnetized Neutron Star (NS) that accretes matter from a low-mass companion ($<$ 1 M$_{\odot}$) through a Roche-Lobe overflow configuration. A distinct framework for comprehending the various accretion states and their variability found in these sources was provided by studying their spectral and timing behavior (Hasinger \& van der Klis 1989). The patterns in X-ray hardness-intensity diagram (HID) or colour-colour diagram (CCD) led to the emergence of two major categories: Z sources and Atoll sources. According to Hasinger \& van der Klis (1989), the Z track pattern found in the HID of Z-sources consists of three branches: the Flaring Branch (FB) at the bottom, the Normal Branch (NB) at the middle, and the horizontal branch (HB) at the top. The mass accretion rate fluctuation over a few days is most likely causing these tracks (Hasinger et al. 1990) and these sources emit close to the Eddington luminosity (L$_{Edd}$ $\sim$ 1). Further investigation ultimately produced two sub-classifications of Z sources: the first is the Sco X-1 like group, which includes the sources Sco X-1, GX 349+2, and GX 17+2, where FB and NB dominate the Z track with a minor contribution from HB; the second group consists of the Cyg-like sources, which include the sources Cygnus X-2, GX 5-1 and GX 340+0, where FB contributes sparsely and rest is dominated by NB and HB in the HID (Kuulkers et al. 1997; Hasinger \& van der Klis 1989).

In the power density spectrum (PDS), quasi-periodic oscillations (QPOs) ranging from a few Hertz to kiloHertz systematically trace the path in the X-ray HID (van Der Klis 2006). In Sco X-1, a linear link exists between kilohertz QPO and normal/flaring branch oscillations (NBOs/FBOs). Because of the high accretion rate in the inner region of the accretion disk, NBOs are somehow linked to radiation-dominated accretion flows (Fortner et al. 1989; Miller and Lamb 1992). However, Alpar et al. (1992) proposed that this could be because of variations in the dynamical properties of the Keplerian rotation of a thick accretion disk. For the occurrence of a 6 Hz NBO in Sco X-1, Titarchuk et al. (2001) supported a model that linked its production mechanism to an acoustic oscillation of a spherical shell region that appears after the inner accretion disk is destroyed as a result of an abrupt increase in radiation pressure in the inner region of the accretion disk. A salient feature observed here is that, with a slight variation in the mass accretion rate, the geometry of the inner region of the accretion flow transforms from a disk-like to a spherical-like structure. Based on the simultaneous radio and X-ray observations, it is found that the NBO + HBO($\sim$6-8 Hz + $\sim$45 Hz) pair, that is, both appearing simultaneously in the PDS( Motta and Fender 2019; GX 5-1; Sriram et al. 2011) is associated with the launch of ultra-relativistic outflow in Sco X-1 (Motta \& Fender 2019). Moreover, a flat-topped noise component in the PDS (i. e. no QPO below $\le$ 100 Hz) was related to the launch of a ballistic-type radio emission. 

Motta \& Fender (2019) reported the strong association of different noise features of PDS with the relativistic outflows in Sco X-1. The ultra-relativistic flow ejection in radio was found to be associated with NBO/NBO+HBO in the X-ray PDS, whereas lobe ejections were connected with flat-topped noise in the PDS. The exact underlying mechanism of the ejection phenomenon is not known (i.e. which physical component is ejected from the accretion disk causing the jets). Another Z source GX 17+2 was also observed simultaneously in both X-ray and radio during the transition of FBO to NBO (Migliari et al. 2007). It was noticed that during the appearance of FBO, the jet is weak, and as the X-ray variability modulates, the NBO feature in the PDS and radio jet becomes steady. Similarly, during the peak of an outburst in the atoll source 1A 1744-361, Ng et al. (2024) noted that a NBO $\sim$ 8 Hz was associated with a transient ejection event in the radio band but the X-ray and radio observations were slightly offset. 
It is still unclear which physical component/components are ejected from the disk to form a jet in Z sources.
 Sco X-1 is the brightest, persistent X-ray source in the sky (Giacconi et al. 1962) and is located at a distance of $\sim$2.8 kpc (Bradshaw et al. 1999) with an inclination angle, found to be $\theta$ = 46$^{\circ}$ $\pm$ 6$^{\circ}$ (Fomalont et al. 2001). It has an M-type companion with a mass of 0.4 M$_{\odot}$ with an orbital period of 18.9 h (Steeghs \& Casares 2002). Sco X-1 is one of the best-studied sources in the radio band among the Z-sources (Andrew \& Purton 1968; Fomalont et al. 2001 a, b).

For the first time, we studied the cross-correlation function (CCF) in seventeen observations of Sco X-1 where simultaneous X-ray and radio observations were performed (Fomalont et al. 2001a, b; Motta \& Fender 2019) and attempted to constrain the accretion-ejection phenomenon and establish an X-ray-Radio association.

\section{Observations and Data Analysis}
The archival data from the {\it Rossi X-ray Timing Explorer} (RXTE) Proportional Counter Array (PCA, Jahoda et al. 2006) of Sco X-1 was used to perform detailed timing and spectral analysis. For this, we reconsidered seventeen observations where simultaneous X-ray and radio observations were performed (Motta \& Fender 2019). The Standard 2 configuration data were used to extract background-subtracted light curves and Proportional Counter Unit 2 (PCU2) data were invoked to extract the spectra as it is best calibrated among all the PCU. The PCA tool, PCARSP version 11.7 was used to obtain the response matrices (for more details see Harikrishna \& Sriram 2022). The updated
background model, the PCA history file spanning the entire
mission, the appropriate response matrix, and the calibrated
database were used to extract the spectra. A systematic error of 1\% was added to the respective spectra. HEASOFT v 6.31 software sub-packages were utilized to reduce the data.
Xspec version 12.13 (Arnaud et al. 1996) was used for spectral analysis and parameter uncertainties were calculated at the
90\% confidence level, i.e. $\Delta \chi^{2}$ = 2.71.

\section{Timing Analysis: Cross-Correlation Function (CCF)}

Energy-dependent light curves in the 2-5 keV (soft band) and 16-30 keV (hard band) were used to perform 
CCF for all the 17 observations in Sco X-1 (see Table 1). Figure 1 displays the hardness intensity diagram (HID) where the intensity was obtained in a 2-20 keV band and hardness is the ratio of $\frac{16-30 keV}{ 2-5 keV }$ bands. Figure 2 shows the soft and hard band light curves (left panels) along with the respective CCFs in the middle panels and the respective power density spectrum (PDS) are displayed in the right panels. It was observed that CCFs are highly correlated, with correlation coefficients (CC) around $\sim 0.8-0.95$ in the NB/FB and the corresponding PDSs are mostly associated with NBO or FBO features. Figure 3 displays the same for the observations where delays were observed in the CCFs. It was noted that a strong asymmetry in CCFs was observed in five observations. Initially, delays were determined by fitting a Gaussian to the peak of the CCFs (Sriram et al. 2012; Sriram et al. 2019). A correlated hard delay of 288$\pm$19 s was seen in the $\dagger$ 5 with correlation coefficient CC = 0.62 (Figure 3, top panel). All the segments of the light curve display correlated and anti-correlated delays varying around a few hundred seconds, i.e. 100 to 346 s. In the third section (c), a correlated hard lag of 103$\pm$22 s was present (second-row panels in Fig. 3). A soft lag of 88$\pm$34 was observed in $\dagger$ 8 with a broad asymmetric CCF.

An anti-correlated hard lag of 207$\pm$28 s was observed in $\dagger$ 9 and a correlated soft lag of  -130$\pm$44 s was observed in the CCF of $\dagger$ 10 (Fig. 3, bottom panels). Notably, these delay-associated observations are linked to the horizontal branch (HB) and hard apex, where a flat-top noise is often noted $\dagger$5, $\dagger$6, $\dagger$7, $\dagger$8, $\dagger$9, $\dagger$10 (see Table 1). 


To validate the delays, we simulated light curves by randomly varying each data point within the standard deviation of the light curve. This procedure was performed independently for soft and hard light curves and repeated 10000 times. The cross-correlation was performed on these simulated soft and hard light curves and their lags were noted and plotted as a histogram (see Figure 4, right panels). A Gaussian was fit to this histogram and its mean value was found to be well in agreement with the observed lags in the CCF. To verify whether the CCFs were significant, the 95\% confidence plots were obtained based on simulated CCFs generated by performing Cross-Correlation analysis on light curves generated using the Timmer-Konig method (Timmer and Konig 1995) with a $\Gamma_{PSD}$ = 2.1, mean and standard deviation of the respective light curves with a bin of 32 s (see Figure 4, left panels). 
The blue-shaded region in the panels displays all the simulated CCFs, and the black dashed line represents the 95\% confidence level. The observed CCF for each ObsIds is overlaying (red line). This indicates that the detected asymmetry in the CCF, and delays were significant and not an outcome of any random process. We also performed the procedure for $\Gamma_{PSD}$ = 1.8 which resulted in a similar result.

\section{Spectral analysis}
We performed spectral analysis of two observations $\dagger$4 and $\dagger$11. In the former, a core flare emission was observed (Motta \& Fender 2019) and the PDS varied from NBO to NoNBO in the PDS (see Figure 5, top left panels). In the case of $\dagger$ 11, another core flare was observed immediately ($\sim$0.5 - 1.5 hr) before this event. In this observation, the PDS changed from NoNBO to NBO (Figure 6. top left panel). The associated CCFs are illustrated in the Results and Discussion section. We took spectra of NoNBO and NBO sections and modeled them. We used a 
 multi-component model, viz. Tbabs*(bbody+Gaussian+nthcomp) where Tbabs represents Galactic
absorption, fixed at N$_{H}$ = 0.3 $\times$ 10$^{22}$ cm$^{-2}$ (Church et al. 2012), and the Gaussian line energy was fixed at 6.4 keV throughout the fitting to model the Iron K emission. {\bf The bbody is a single temperature black body model for describing the X-ray emission from the boundary layer (BL) and nthcomp is a model for the thermal Comptonization that accounts for the hard coronal emission which characterizes the corona in the inner region of the accretion disk (Zdziarski et al. 1996).} 
The bbody model parameters are: black body temperature kT$_{bb}$ and normalization N$_{bb}$. nthcomp has the following parameters: asymptotic power-law photon index ($\Gamma_{nthcomp}$), the electron temperature (kT$_{e}$), and the seed photon temperature (kT$_{soft}$) along with normalization (N$_{nthcomp}$).


In case of $\dagger$4, the black body temperature was found to be kT$_{bb}$ $\sim$ 1.35 keV along with a $\Gamma_{nthcomp}$ $\sim$ 2.0, kT$_{e}$ $\sim$ 2.87 keV, and kT$_{soft}$ $\sim$ 0.30 keV. We did not notice any variation among spectral parameters and a similar result was noted in $\dagger$ 11 (see Table 2). Similar values were observed for Sco X-1 using the Insight-HXMT satellite's LE+ME instrument's spectra spanning 2-35 keV (Ding et al. 2023).  We also computed the unabsorbed flux in the 3.0 – 25.0 keV energy band using the convolution model cflux along with additive models. Among the NBO and NoNBO spectra, it was observed that the fractional flux of the bbody component i.e. $(\Delta f/f)_{bb}$ varied by $\sim$ 10-20 \% in two observations (see Table 2 and Figure 7). In contrast, $(\Delta f/f)_{nthcomp}$, the fractional flux of the nthcomp component remained similar.

\section{Results and Discussion}
We examined the cross-correlation functions between the soft (2-5 keV) and hard (16-30 keV) energy bands of Sco X-1 for seventeen observations linked to concurrent X-ray and radio monitoring (Motta \& Fender 2019; Fomalont et al. 2001 a, b). We found that CCFs became highly symmetric during the appearance of NBO or NBO+HBO in the respective PDS across the NB (Fig. 2, Table 1) and {\bf the estimated centroid frequencies of the QPOs are well compatible with the ones obtained by Motta \& Fender (2019) for the same data}. When the CCFs became asymmetric exhibiting correlated/anti-correlated delays, the PDS was dominated by a flat-topped noise (see Fig. 3) in the horizontal branch (HB) and hard apex. During $\dagger$5, a flat-topped noise was observed after the radio lobe ejection (see Table 1). Before this observation, a broad NBO feature was observed in $\dagger$4 where the CCF was highly correlated, displaying no delays (Fig. 5, top left panel). During the inspection of the light curve ($\dagger$4), we noted that in the initial section, the CCF was symmetric with no delay (Fig. 5, top right panel) exhibiting an NBO in the PDS (Fig. 5 right bottom panel, cyan line). In the later section of the light curve, the NBO disappeared but a flat-topped noise was seen in the PDS (Fig. 5, bottom right panel, green line) which was connected to an asymmetric and complex CCF with a possible soft delay (Fig. 5 bottom left panel). Radio observations unveiled a core flare emission (C1) close to this observation (see Table 1). 

It is interesting to note that in $\dagger$5 and $\dagger$4 with NoNBO, exhibit a similar flat-topped noise in the PDS (see Figure 5, right bottom panel, black line ($\dagger$5) and cyan $\dagger$4). After 40 min of $\dagger$5 (ObsId. 40706-02-09-00), a core flare emission (C2) (Motta \& Fender 2019) was seen in the radio band which was later accompanied by flat-topped noise in the respective PDS, and delays were detected in the CCFs of $\dagger$5, $\dagger$7, $\dagger$8, $\dagger$9 and $\dagger$10 (see Fig. 3) This suggests that CCFs display remarkable and systematic variations as the Sco X-1 moves across horizontal and normal branches. Around the same time, in $\dagger$10 (ObsId. 40706-02-14-00), a core flare (C3) was observed accompanied by lobe flares and later within 45 min, an NBO+HBO pair appeared in $\dagger$11. Again in this observation, the CCF was highly correlated displaying no delay (Fig. 6, top left panel) in the initial phase of the light curve up to 2200 s, a flat-topped noise component was seen in the PDS (Fig. 6 bottom right, black line) with a slight asymmetric CCF with a small soft delay of -90$\pm$32 (Fig. 6, top right panel). Later an NBO appeared (Fig. 6, bottom right panel, cyan) associated with a narrower CCF without delay (Fig. 6, bottom left panel). A similar flat-topped noise in the PDS of $\dagger$11 and $\dagger$10 is again observed. Overall based on the above discussion, we propose that during the flat-topped noise where delays were detected in CCF or the CCF became asymmetric, the inner region of the accretion disk experiences instability due to the triggering of ballistic-type jet, and delays are nothing but the readjustment time scale of this region.  

\textbf{During the NBO to NoNBO variation, we did not find any variation in the spectral parameters but noticed that the fractional flux variation in the bbody component, i.e. ($\Delta$ f/f)$_{bb}$ was higher than the fractional flux variation of nthcomp ($\Delta$ f/f)$_{nthcomp}$ (see Table 2).
In ObsId. 40706-01-01-00, ($\Delta$ f/f)$_{bb}$ was $\sim$ 10\%, while for nthcomp, it was found to be less, ($\Delta$ f/f)$_{nthcomp}$ $\sim$ 4\%. A similar fractional variation was seen in ObdsId. 40706-01-02-00, ($\Delta$ f/f)$_{bb}$ $\sim$ 20\% and ($\Delta$ f/f)$_{nthcomp}$ $\sim$ 1\%.} We conclude that, since the fractional variation of the bbody flux is higher than the fractional variation of the nthcomp fractional flux, probably the boundary layer is readjusted during these observations and is responsible for observed delays and asymmetric CCFs. It is interesting to note that for Sco X-1, Jia et al. (2023) also reported a small variation in the total flux ($\sim$ 12 \%) between the NBO and NoNBO epoch spectra using NICER data. Moreover, based on NICER and NuStar data, a recent study also found that CCF was symmetric exhibiting no delay when the Z source Cyg X-2 was in the normal branch (Malu et al. 2024).

Motta \& Fender (2019) associated two different types of radio emissions, viz. ballistic ejections (lobe ejections) associated with the HB/hard apex and ultra-relativistic flow (URF) ejection associated with the NB. We noted that NBO or NBO+HBO in the PDS is often accompanied by highly symmetric CCFs with no delays (see Figure 2). The highly symmetric CCF possibly indicates that the inner region's physical components of the accretion disk responsible for soft and hard emissions are in a stable configuration. Moreover, the radio spectral index of lobe ejection and URF does not differ substantially (Fomalont et al. 2001a) and hence it is difficult to determine whether these different radio ejection events are connected to the same or different physical scenarios in the accretion disk. Since our investigation reveals that the CCFs are fairly symmetric, we hypothesize that the URF linked to NB is probably a delayed lobe emission (see also Motta and Fender 2019). High-resolution radio and X-ray observations are required to confirm this scenario.


The exact production mechanism of NBO is not clearly understood,  but it may arise due to oscillation in the optical depth of the radial inflow (Lamb 1989; Fortner et al. 1989) or it may also be associated with the 
 spherical shell oscillation in the inner region of the accretion disk (Titarchuk et al. 2001). Using the NICER data, Jia et al. (2003) suggested that NBO in Cyg X-2 and Sco X-1 could be linked to dual Comptonization in the transition layer structure influenced by accretion disk soft photons below $<$1.5 keV and neutron star photons above $>$1.5 keV. 
 In another source, Cyg X-2, NBO was detected with NICER satellite data with peak modulation occurring at 1-2 keV (Malu et al. 2024). Invoking a broad-band spectrum using NICER + NuStar data, it was noted that both kT$_{bb}$ and kT$_{in}$ between the NBO and NoNBO epochs varied and the possible changes were connected to the boundary layer (BL) and inner accretion disk regions (Malu et al. 2024). However, the origin of NBO to any specific location was not concluded. Interestingly, NBO was associated with an outer region of the accretion disk ($\sim$ 300 km from NS) due to the peaking of the NBO feature in the 1-2 keV energy band, its absence in the $>$ 3 keV energy band, and the association of a Fe L line in the corresponding energy spectra (Malu et al. 2024). Our study found that NBO or NBO+HBO were linked to highly symmetric CCFs where URFs were often detected in the radio band. Interestingly, with the NICER and NuStar energy bands, Malu et al. (2024) also reported symmetric CCFs with no delays during NBO observation indicating a stable accretion flow with a compact corona structure. In general, it is challenging to determine which part of the inner accretion disk is responsible for the jets (URF), i. e. is it the compact corona modulating HBO, or is it the transition/boundary layer, or an outer region of the accretion disk causing the NBO.

 Radio observations indicated that the jet is ballistic-type at the hard apex as seen in Sco X-1. The delays and asymmetry of the CCF suggest that the inner region of the accretion disk is unstable, which explains the absence of oscillation features, viz. HBO/NBO in the PDS. The delays can be attributed to the readjustment time scale of the corona or boundary layer as discussed in other Z sources (Sriram et al. 2012; Sriram et al. 2019; Malu et al. 2020; Chiranjeevi et al. 2022). The cause of instability is unknown, but it could be triggered due to high radiation pressure driven by an increase in the mass accretion rate (eg. Church, Gibiec \& Bałucinska-Church 2014). A jet might emerge due to this phenomenon, which could cause the material in the inner accretion disk to be launched vertically.



\subsection{CCFs of other Z sources}

Migliari et al. (2007) reported a strong association between NBO $\leftrightarrow$ FBO transition and jet ejection phenomena in a Z source, GX 17+2, using simultaneous X-ray and radio observations. During these observations, Sriram et al. (2019) reported a correlated hard delay of 100$\pm$28 s (associated with a strong radio flux), which was readily seen in the first portion of an observation with ObsId. 700023-01-01-01 exhibiting an HBO of $\sim$ 25 Hz. This is followed by an anti-correlated CCF along with a soft lag of -168 $\pm$ 42 s associated with low radio flux. These delays were interpreted as the readjustment time scale of the corona, which emits hard photons in the inner disk region (Sriram et al. 2019). In GX 17+2, it was proposed that if the hard photons come from the base of the jet, then the hard flux tends to decrease with the shrinkage of the jet base and delays were connected to the readjustment time scale of the jet's base. In a different ObsId. 70023-01-02-00 of GX 17+2, the radio flux was low and the X-ray variability began without any noise characteristic in the PDS. Over the next 300 s, the radio flux steadily grew, accompanied first by an FBO and then by an NBO. During this event, radio flux suddenly increased and the jet stabilized when the FBO transformed to an NBO (see Fig. 3 in Migliari et al. 2007). Using AstroSat data, similar X-ray delays between 3-5 keV, 0.8-2.0 keV (soft bands), and 10-20 keV, 16-30 keV (hard bands) in GX 17+2 were later detected (Malu et al. 2020; Sriram et al. 2021) which were interpreted as the readjustment time scale of the corona. Such delays were also seen in GX 5-1 (Sriram et al. 2012) and Cyg X-2 (Lei et al. 2008) densely occupying around hard apex and HB region of HID. The delays in CCFs of a few tens to hundreds of seconds are ubiquitous in Z sources. From the present study, {\bf \it we suggest that asymmetric CCFs with delays are strong evidence that the inner region of the accretion disk experienced instability due to the triggering of a ballistic jet in Sco X-1, and hence, only flat-top noise was present with no oscillation features in the PDS.}

\subsection{Possible Association of Delays with Boundary Layer}
The delays observed between the hard and soft light curves from their cross-correlation analysis could be attributed to the depletion time scale of the boundary layer in the neutron star. Matter falling onto the NS from the accretion disk cannot directly reach the NS surface as they both have a substantial difference in angular momentum, and the matter in the accretion disk orbiting at Keplerian frequency needs to slow down before touching the NS surface to match its rotational frequency. Hence, a layer forms over the NS surface, often known as the boundary layer (for more details, see Popham \& Sunyaev 2001).
It was demonstrated that the formation of BL is transient, and two different time scales are often associated, i. e. depletion and frictional time scales (t$_{depl}$ and t$_{fric}$; Abolmasov et al. 2020; Abolmasov \& Poutanen 2021). Both time scales were found to be higher than the viscous time scales in this region. The frictional time scale is a duration where the BL and NS angular velocity attain the equilibrium ($\Omega$=$\Omega_{NS}$), and the depletion time scale is a replenishing or depleting duration of the material in the BL. The equation $t_{depl}/t_{fric} = \alpha \Omega_{K} t_{depl}$ connects these time scales. Here, $\Omega_{K}$ represents the Keplerian frequency, and $\alpha$ is a proportionality constant between the stress and pressure terms, ranging from 10$^{-6}$ to 10$^{-7}$ (refer to equation 2 of Abolmasov \& Poutanen 2021). In an atoll source 4U 1728-34, the depletion time scale was determined to be t$_{depl}$ = 740 s for $\alpha$ = 10$^{-7}$ and smaller for $\alpha$ = 10$^{-6}$ (Chiranjeevi et al. 2023). 


The depletion time scale depends on the rotation frequency as 
t$_{depl}$=$\frac{(\Omega - \Omega_{NS})}{\alpha (\Omega_{k}^2-\Omega^2)}$, where $\Omega_K = \frac{1}{2\pi}\sqrt{GM_{NS}/R^3}$  and $\Omega_{NS}$ = 649 Hz (Yin et al. 2007).  For R$_{NS}$ = 10 km of neutron star radius and M=1.4 M$_{\odot}$, $\Omega_K$  = 2169.77 Hz, $\Omega$ = 767.13 Hz for the material orbiting at a radius of R = 20 km, we get $t_{depl}$ = 28.67 s ($\alpha = 10^{-6}$) and $t_{depl}$ = 286.67 s ($\alpha = 10^{-7}$). Similarly for M=2 M$_{\odot}$, $t_{depl}$ = 45.52 s ($\alpha = 10^{-6}$) and $t_{depl}$ = 455.22 s ($\alpha = 10^{-7}$). {\bf However, it should be noted that $\Omega_{NS}$ = 649 Hz is an upper limit, and a lower value would result in a higher depletion time scale.} The calculated depletion time scales are comparable to observed delays in the respective CCF. Recently, Chiranjeevi et al. (2023) showed that delays observed in an atoll source 4U 1728-34 could be described as a result of depletion time scale in the boundary layer using AstroSat data, and similar delays were seen in another atoll source 4U 1705-44 (Malu et al. 2021).  Comprehensively, we suggest that the detected delays are plausible readjustment time scales of the boundary layer. A stable jet configuration cannot be achieved, and the oscillation features in the PDS vanish because the jet base cannot form while the BL is readjusting to achieve a stable configuration. 

\textbf{It is yet unclear which spectral component is experiencing more variation during the CCF delay. The observations associated with delays occupied HB or hard apex in the HID for Sco X-1 (see Figure 1) with a slight variation in hardness, indicating a low flux variation.  Hence, a direct connection to the spectral flux variation of the BL depletion scenario is difficult to establish. A variety of emission scenarios originating from various physical components of the accretion disk can be used to model the X-ray spectra of Sco X-1. \\
Church et al. (2012) proposed an accretion disk corona (ADC) model for Sco X-1 and other Z sources. It was found that the blackbody temperature, kT$_{bb}$ was $>$ 2 keV in the soft apex and Comptonized emission component dominates the overall luminosity while the blackbody component was contributing at a level of 10\% only. Moreover, there is a steep increase in the Comptonized emission at NB as the source traverses from hard apex to soft apex (Church et al. 2012). In this scenario, the Compotonized component has a slab-like structure over the accretion disk and the delays in CCF would suggest that an overall slab-like corona would have been disrupting the entire structure. Such a scenario has a low possibility as the ADC structure is often found to be located at radial positions between 18000 and 110000 km (eg. Schulz et al. 2009) in Cyg X-2. For LMXBs, Church \& Bałucińska-Church (2004) also constrained the size of the ADC to be around $\sim$20000 km to 700000 km depending on the luminosity of the source. In another physical scenario for Sco X-1, Titarchuck, Seifina \&  Shrader (2014) proposed a physical structure where the transition layer’s (TL) outer region is comptonized by the disk photons, and the inner region of TL is comptonized by the soft photons arriving from the neutron star surface. Since TL is close to the NS, the disturbances of the TL would result in assymmetric 
 CCFs. While this may potentially account for the lack of NBO in PDS and observed asymmetry in CCF, more investigation is necessary to establish a direct link between the radio ejection and TL structure in the inner region of the accretion disk.}


It is still unclear exactly how the jet is turned on, however, it most likely happens near horizontal and normal branches (e.g. Migliari et al. 2007; Motta and Fender 2019). Based on accretion-ejection models, the presence of corona plays a pivotal role in the jet phenomenon. A hot, thin, and extended corona does help in the formation and ejection of a jet eg. (Mendez et al. 2022 for a BH source GRS 1915+105). Such an extended corona model was already being explored in a Z source Cyg X-2 (Bałucinska-Church et al. 2010). Both disk and NS/BL photons control the geometrical thickness and temperature of the corona. During the low accretion phase, the corona is relatively large due to low soft photon influx, making it extended, which is needed for material ejection. In the case of Sco X-1, the detected delays suggest that the inner region, possibly the BL along with the corona, is readjusting due to instability caused by the ballistic jet. During the readjustment time, the onset of instability in this region quenches the launch of the jet. Though direct evidence is lacking in Sco X-1, however, such a BL-related jet probably seems to be occurring in the atoll source 4U 1820-30 (Marino et al. 2023) and in a dwarf nova SS Cyg where jet launching was found to be possibly connected with the formation of the boundary layer (Russell et al. 2016).

\subsection{Possible Region Size of Jet Triggering}
 Fomalont et al. (2001a, b) found that the radio lobes' velocity in Sco X-1 was $\beta$ = {\it v/c} = 0.32-0.57 over a few hours.
 Assuming this to be the escape velocity, the minimum size of the jet emitting region should be $\sim$ 12-40 km, which indicates that the transient jet is triggered close to the neutron star, possibly causing the observed asymmetry in the CCF. Titarchuk et al. (2001) proposed that NBO/FBO is caused due to the oscillation in a spherical shell around the neutron star's surface. The shell viscous frequency is given by $\nu_{NBO}$ = $\frac{f V_{s}}{L}$, where f ranges from 0.5--$\frac{1}{2\pi}$, $V_{s}$ is the sonic velocity that is estimated based on Equation \begin{equation}
     v_{s} = 4.2 \times 10^{7} R_{6}^{-1/4} (\frac{M}{M_{\odot}} \frac{L}{L_{\odot}})^{-1/8} cm s^{-1}
 \end{equation}, where R$_{6}$
 is the radius of the neutron star in units of 
10$^{6}$cm. For calculations, we consider a radius of 10 km and a mass of 1.4 M$_{\odot}$ for the neutron star and estimated L = 21 km for 10 Hz and 32 km for 6 Hz oscillations. 

\section{Conclusions}
For Sco X-1, we performed the cross-correlation function studies using soft and hard band energy-dependent X-ray lightcurves for seventeen observations where a relativistic jet was present based on previous radio observations. On HB and hard apex, hard and soft correlated/anti-correlated delays of a few tens of seconds were detected, which were confirmed by simulations. We conclude the following from the present study: \\

1. The asymmetric CCFs with delays occurring around HB and hard apex are closely associated with flat-topped noise with no oscillation features in the PDS where radio studies suggest that the jet is ballistic type. 

2. The symmetric CCFs with no delay are found to be associated with NB and exhibit an NBO or NBO+HBO in the PDS. Based on the radio studies in this branch, radio emission of an ultra-relativistic flow (URF) type is observed.

3. We did not find noticeable spectral variation during the NBO to NoNBO transition in two observations where a core flare emission occurred close to the two observations. But both CCFs and PDSs varied abruptly. However, the bbody flux varied by 10-20\%, although such small variations were noticed with NICER studies for Sco X-1 (Jia et al. 2023).

4. The instability in the inner region of the accretion disk is triggered possibly due to a ballistic jet in Sco X-1, resulting in asymmetric CCF with delays and the absence of oscillation features in the PDS. The asymmetry in CCF could be due to sudden disturbances in the soft (possibly the inner disk front) and hard emitting (corona/boundary layer/transition layer) structures in the inner region of the accretion disk caused by the ballistic jet. However, further research is needed to determine what triggers the jet's launch.

5. During the symmetric CCF (without delay), the accretion flow was steady, therefore, the NBO / NBO + HBO features appeared consistently in the PDS when Sco X-1 was in the NB of the HID. Moreover, we speculate that the driving mechanism generating the NBO is also responsible for the URF in the radio band. Since NBO has been found in both Z and atoll sources (Malu et al. 2021; Ng et al. 2024; Wijnands et al. 1999), a high accretion rate could not be a stringent requirement for the production mechanism. As a result, URF may need a phenomenon that is partially reliant on or completely independent of high accretion states.

Our study suggests a strong association between CCFs, PDS, and radio ejection events. Future broad-band X-ray timing and spectral studies using XSPECT onboard XPoSat, NICER, and NuSTAR would be essential to constrain the different mechanisms to understand the accretion-ejection phenomenon in Sco X-1 and other Z-type sources.

\section{Acknowledgments}

We thank the Referee for providing comments which improved the quality of the paper. K.S and SD acknowledge financial support from the ANRF Core Research Grant project, the Government of India. This research has used data and or software provided by the High Energy Astrophysics Science Archive Research Center (HEASARC), a service of the Astrophysics Science Division at NASA/GSFC.

\newpage
\clearpage
\begin{figure}

\includegraphics[height=15cm, width=17.5cm, angle=0]{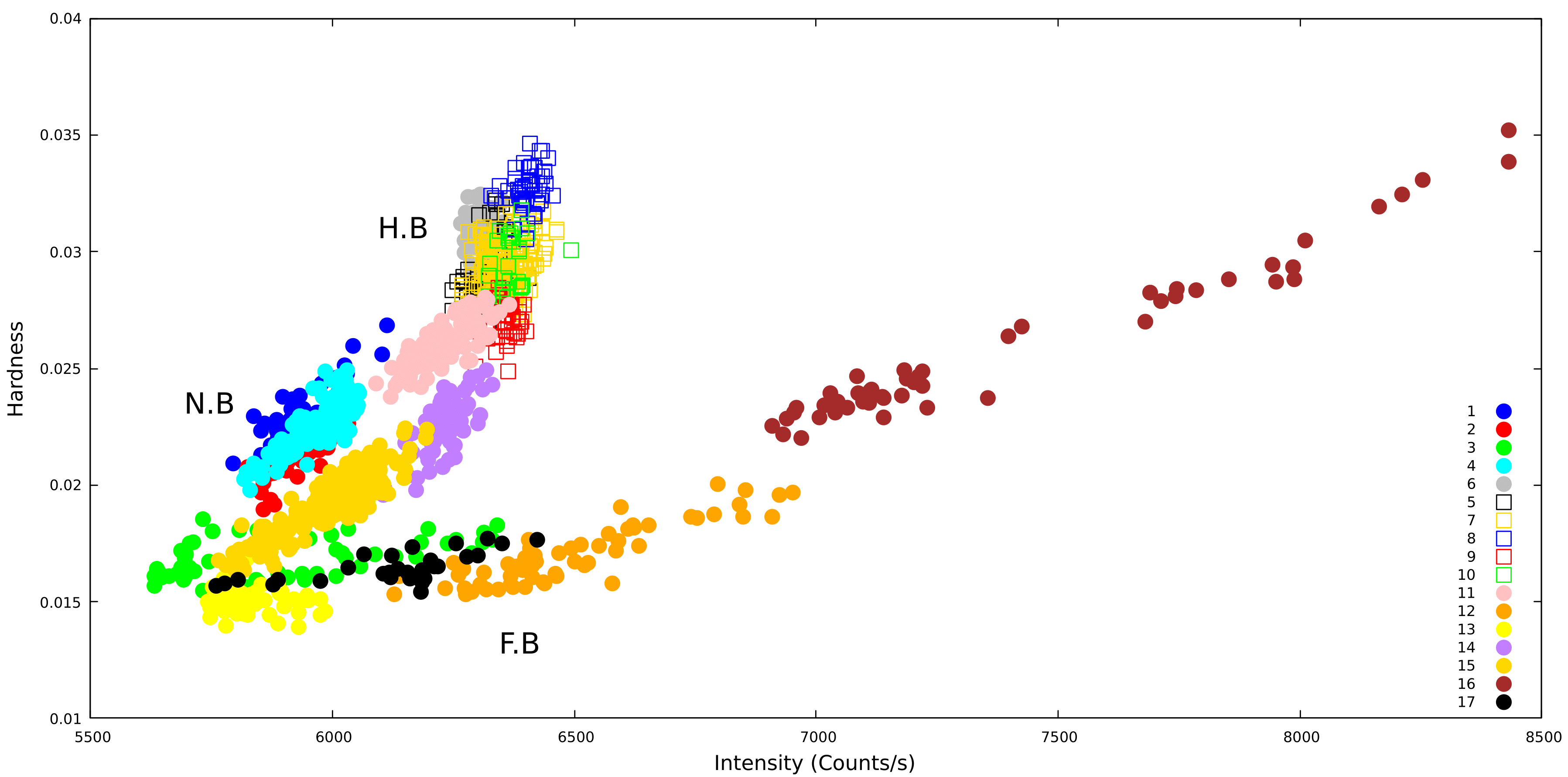} 
\begin{minipage}{\textwidth}
\caption{The Z shape can be depicted from HID for Sco X-1 using the seventeen observations. The hardness was obtained from the ratio of $\frac{16-30 keV}{2-5 keV}$. The delays were detected in the horizontal branch displayed with square boxes. The hard apex is located between HB and NB. }
\end{minipage}
\end{figure}

\clearpage
\begin{figure*}

\includegraphics[height=7 cm, width=17 cm, angle=0]{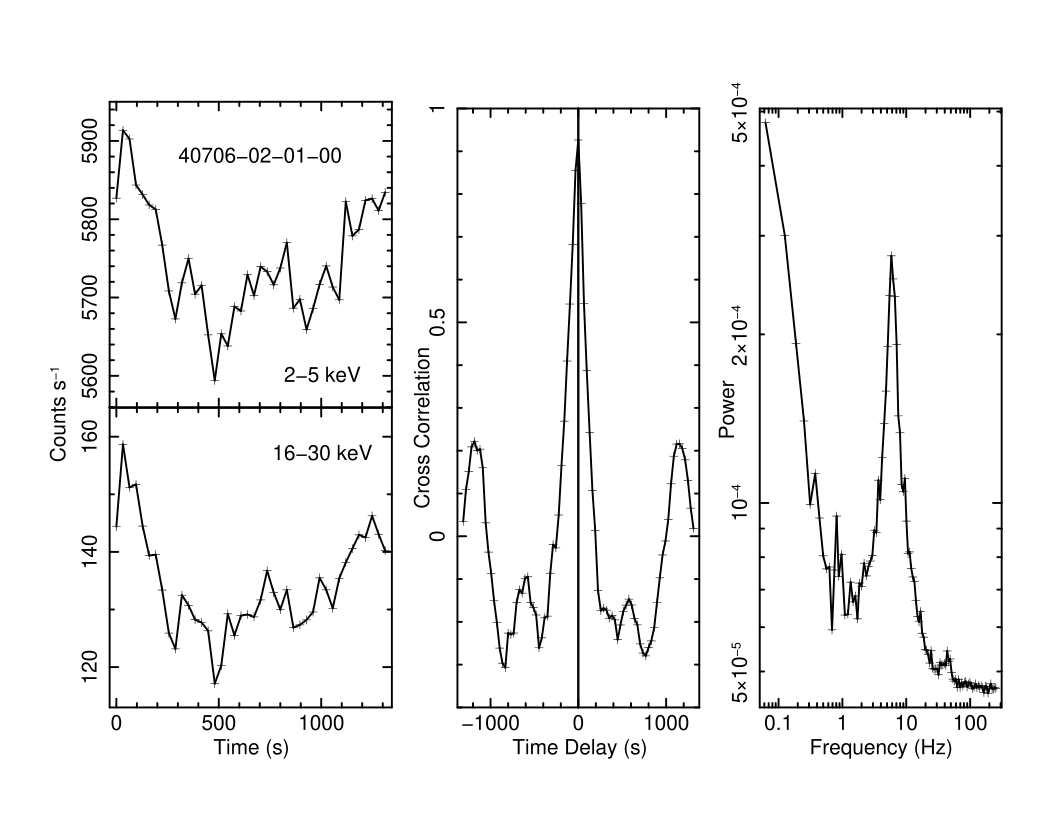}\\

\includegraphics[height=7 cm, width=17 cm, angle=0]{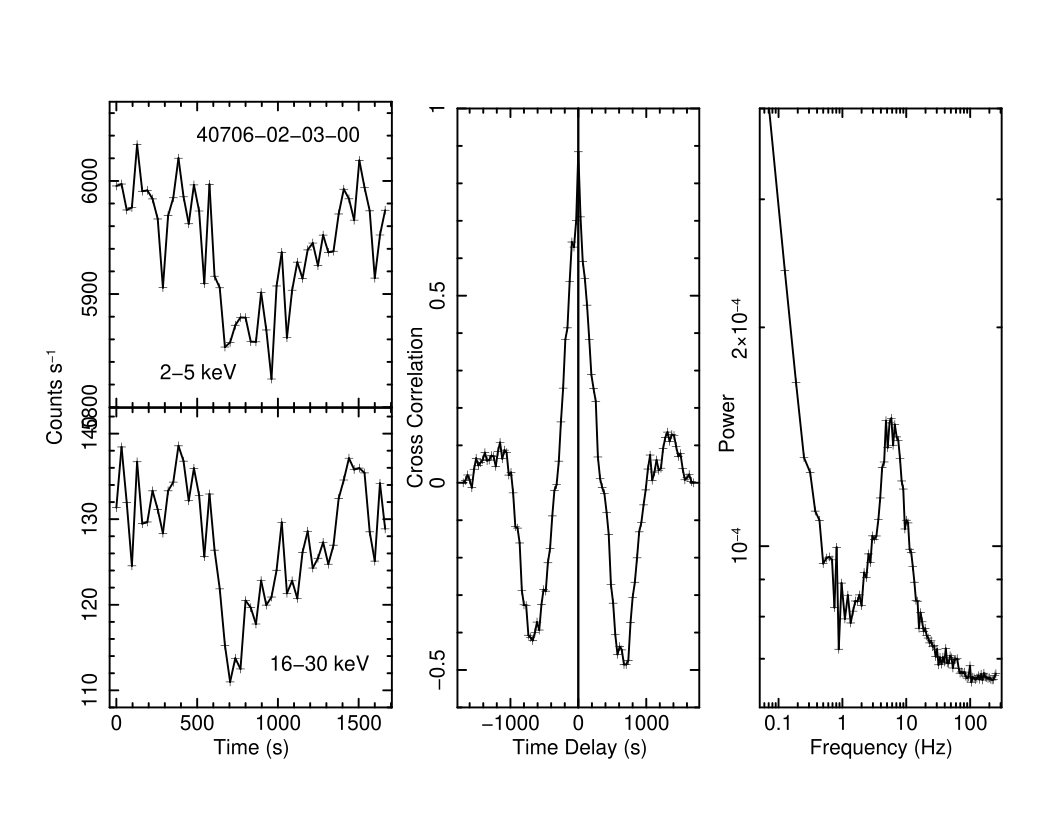}\\ 

\includegraphics[height=7 cm, width=17 cm, angle=0]{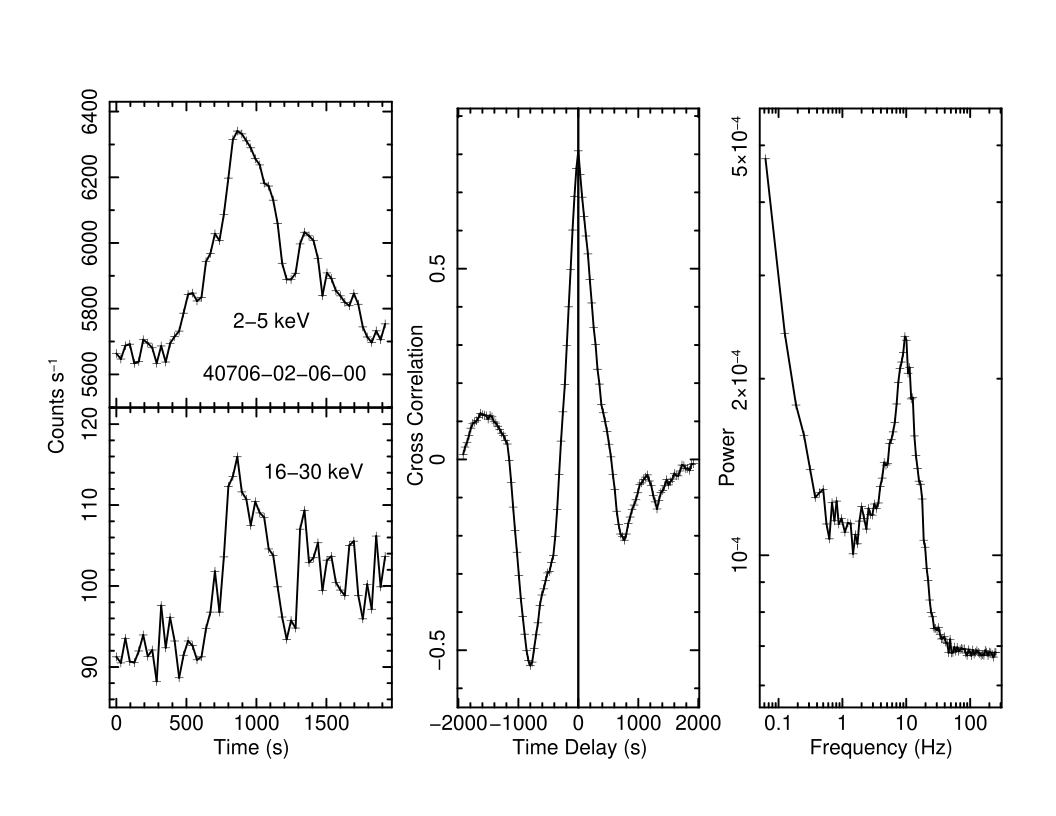}\\

\caption{Displaying the 2-5 keV and 16-30 keV X-ray light curves (left panels), cross-correlation function (CCF, middle panels) where the vertical line shows the zero delay and respective PDS in right panels. The ObsIds are shown in the inset (left panels). }
\end{figure*}

\clearpage
\begin{figure*}
\includegraphics[height=7 cm, width=17 cm, angle=0]{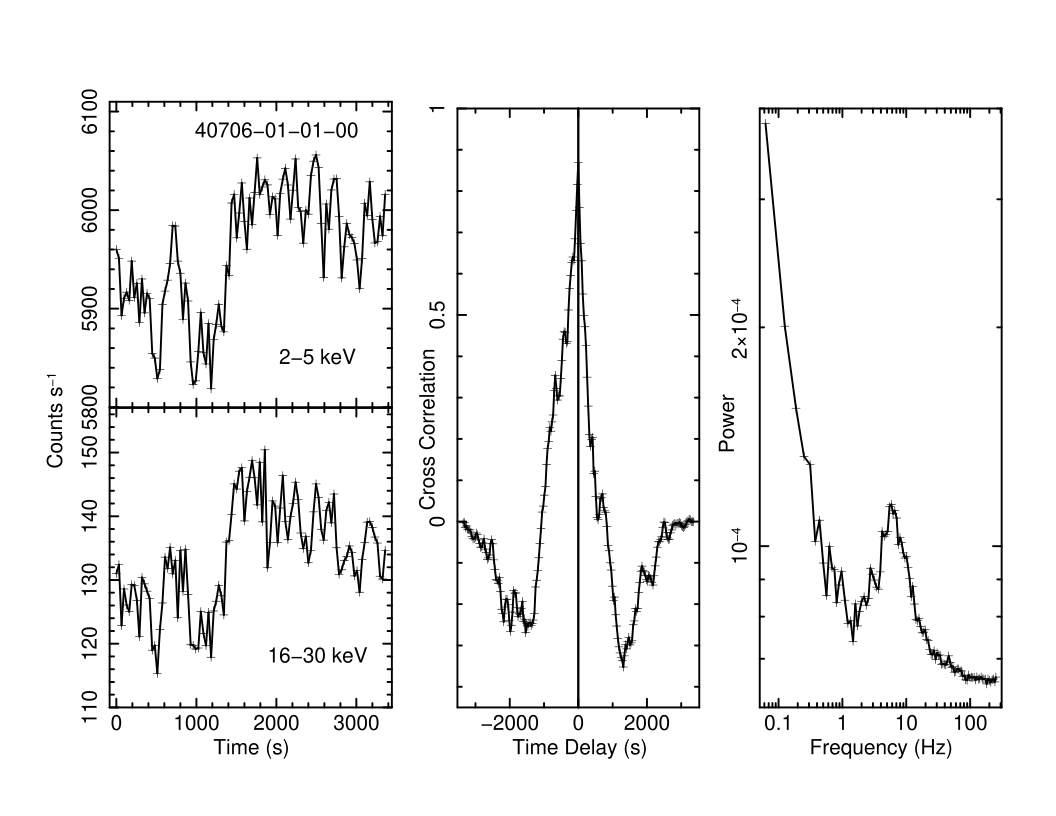} \\ 

\includegraphics[height=7 cm, width=17 cm, angle=0]{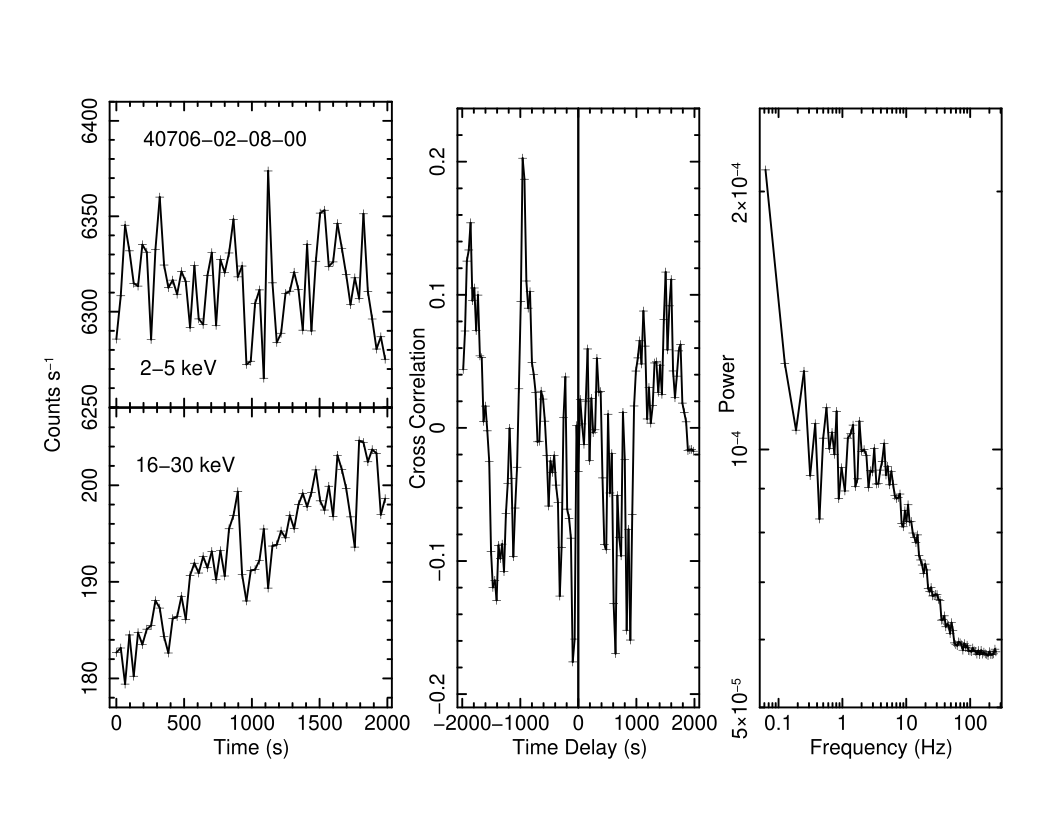} \\

\includegraphics[height=7 cm, width=17 cm, angle=0]{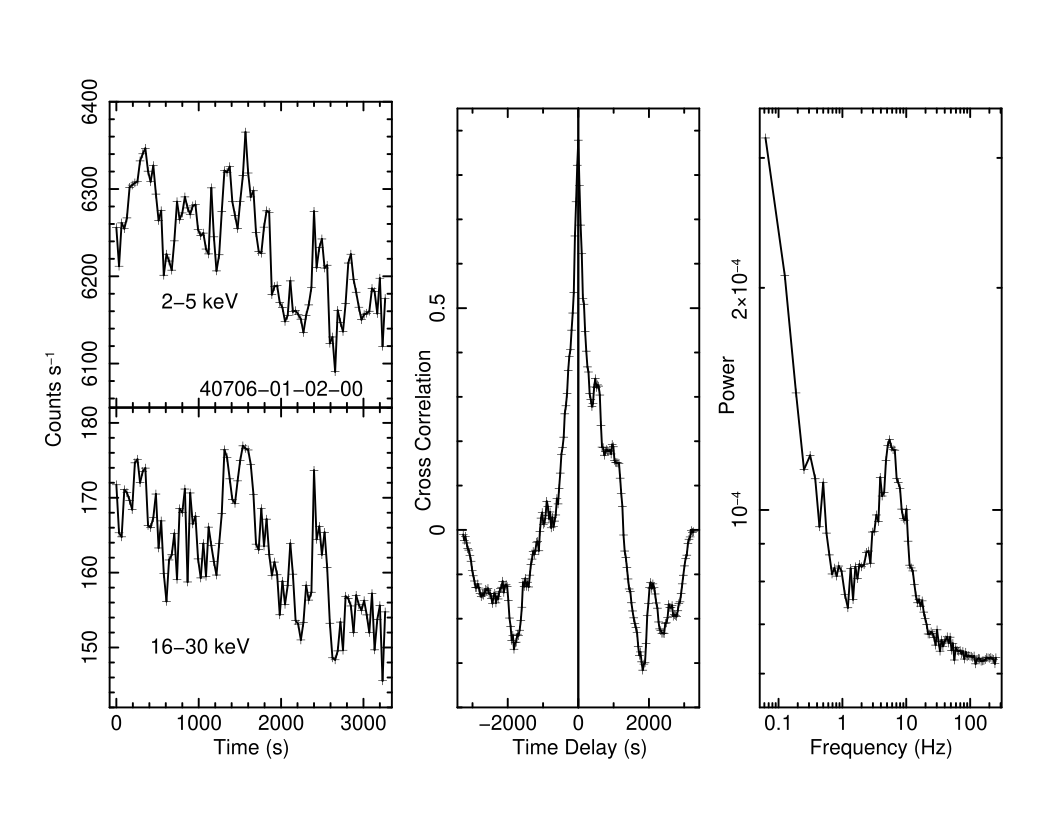}\\


{{\bf Fig. 2}:            Continue.. }


\end{figure*}

\clearpage
\begin{figure*}

\includegraphics[height=7 cm, width=17 cm, angle=0]{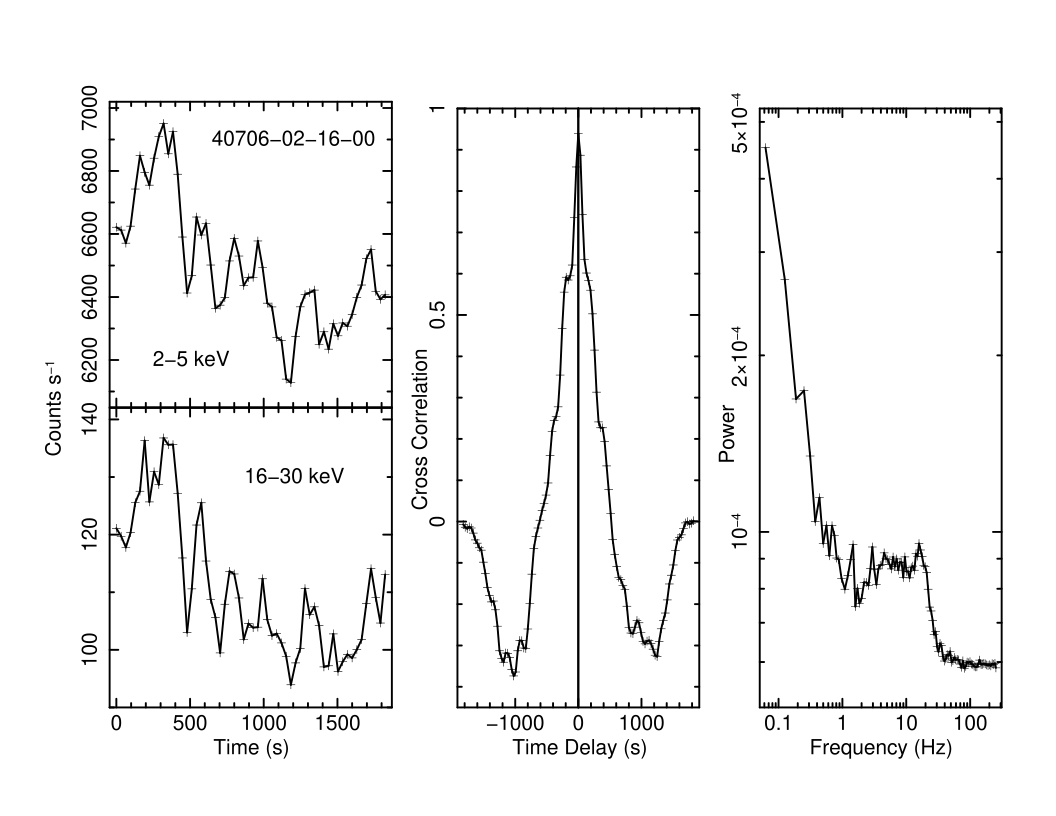}\\

\includegraphics[height=7 cm, width=17 cm, angle=0]{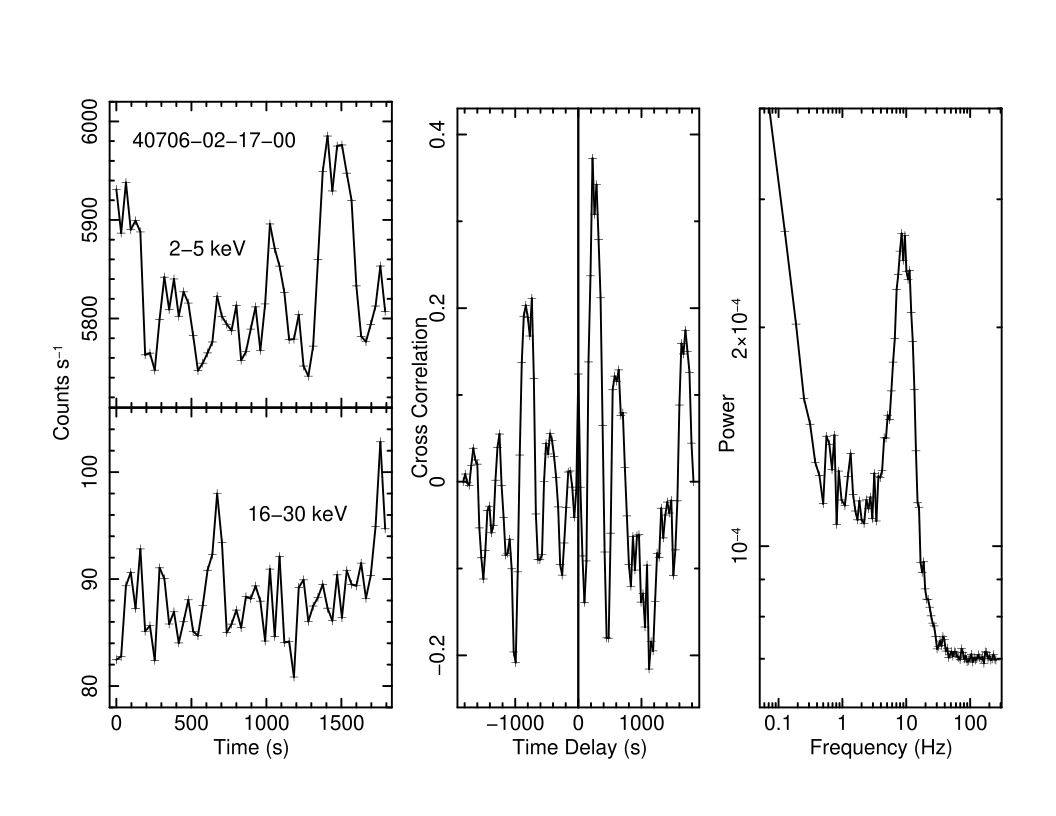}\\

\includegraphics[height=7 cm, width=17 cm, angle=0]{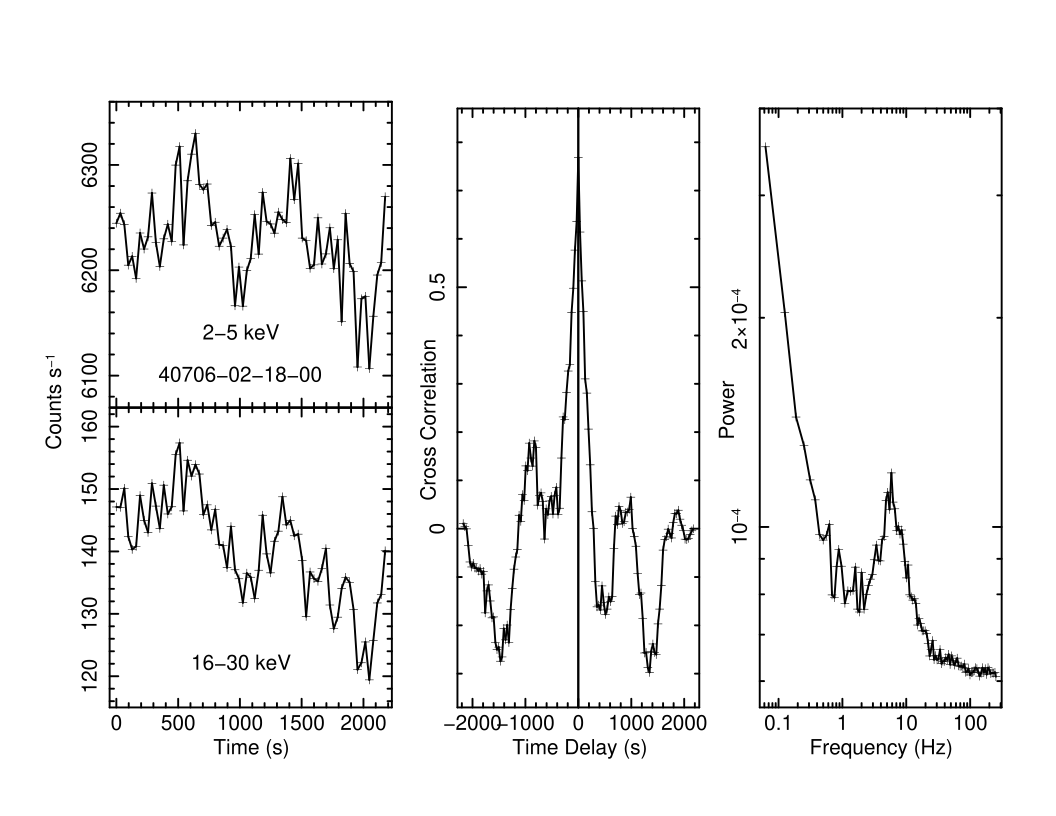}\\

{{\bf Fig. 2}:            Continue.. }
\end{figure*}

\clearpage
\begin{figure*}
\includegraphics[height=7 cm, width=17 cm, angle=0]{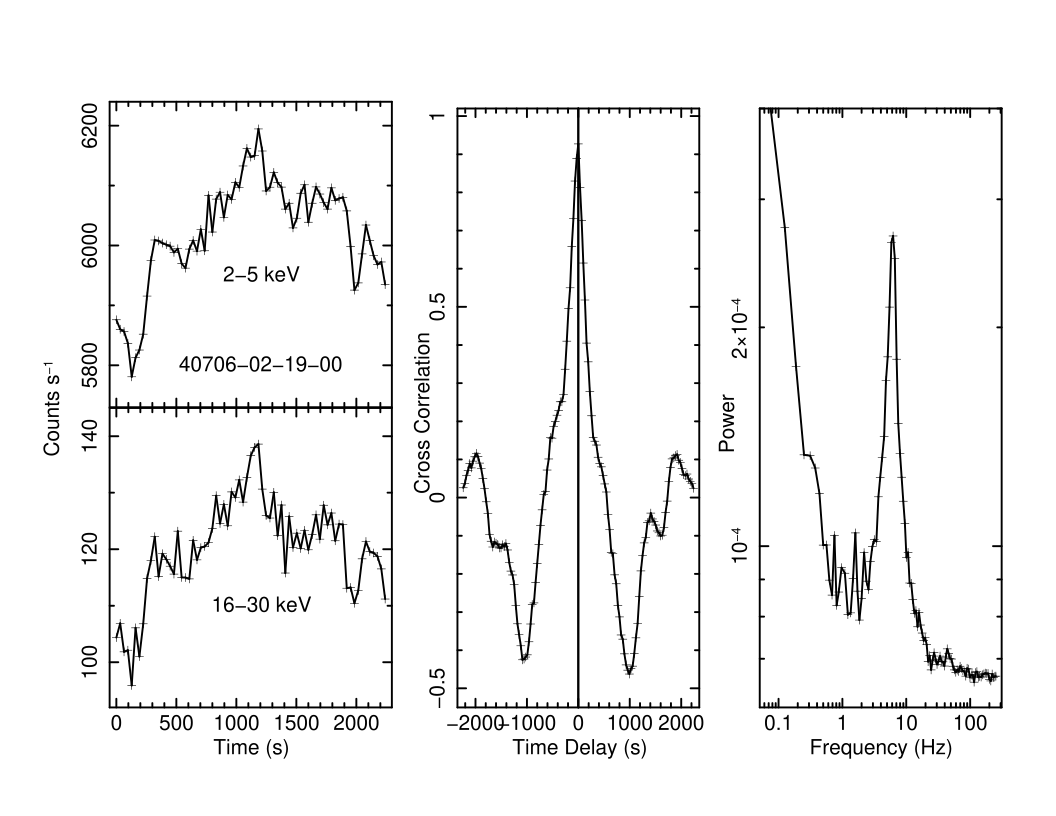}\\

\includegraphics[height=7 cm, width=17 cm, angle=0]{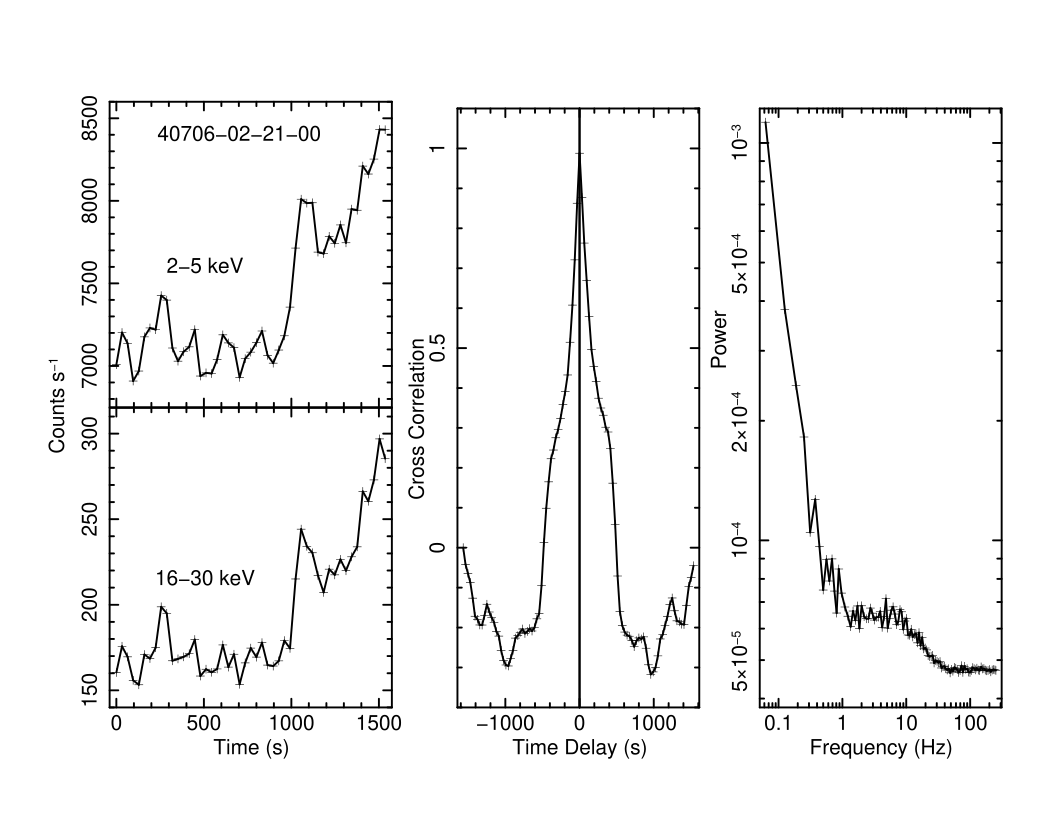}\\

\includegraphics[height=7 cm, width=17 cm, angle=0]{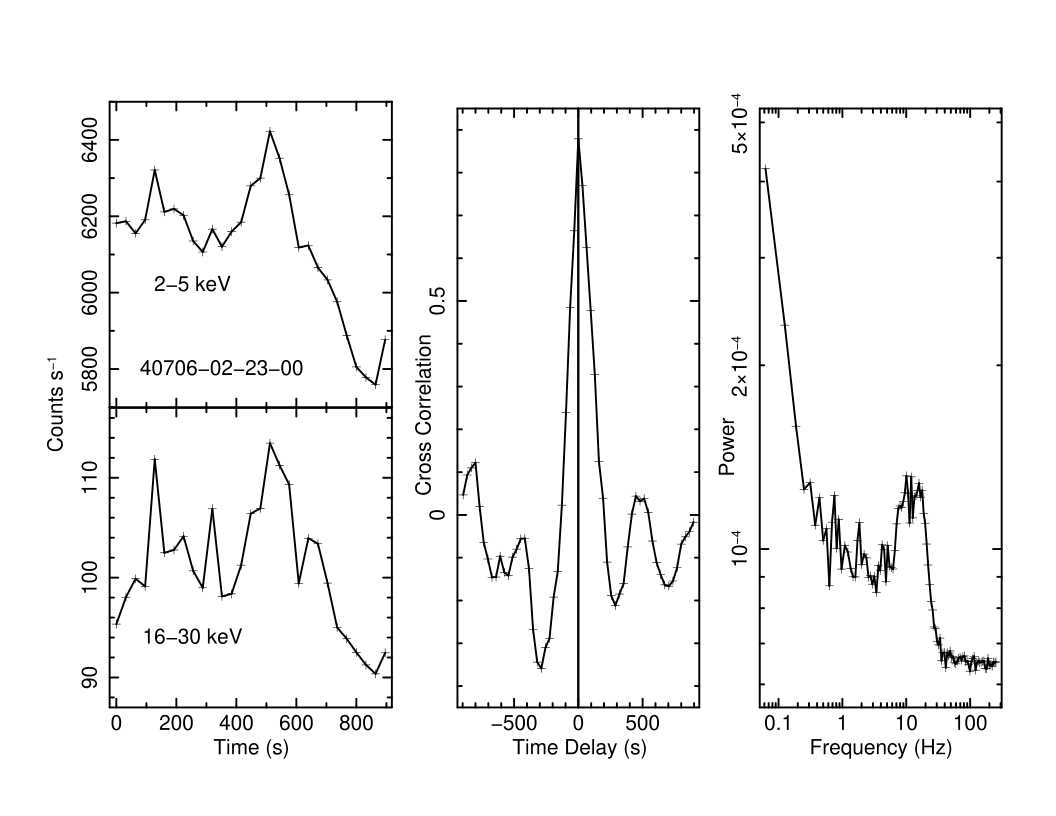} \\ 
{{\bf Fig. 2}:             Continue.. }
\end{figure*}



\clearpage
\begin{figure*}
\includegraphics[height=7 cm, width=17 cm, angle=0]{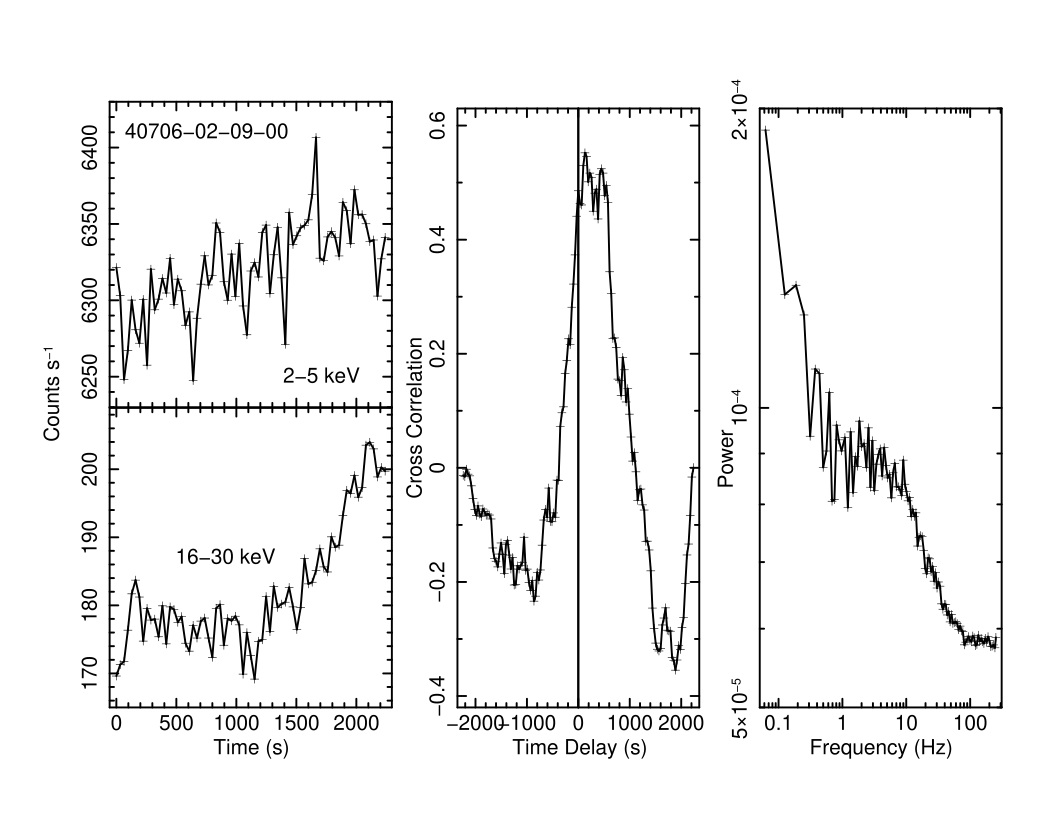} \\ 

\includegraphics[height=7 cm, width=17 cm, angle=0]{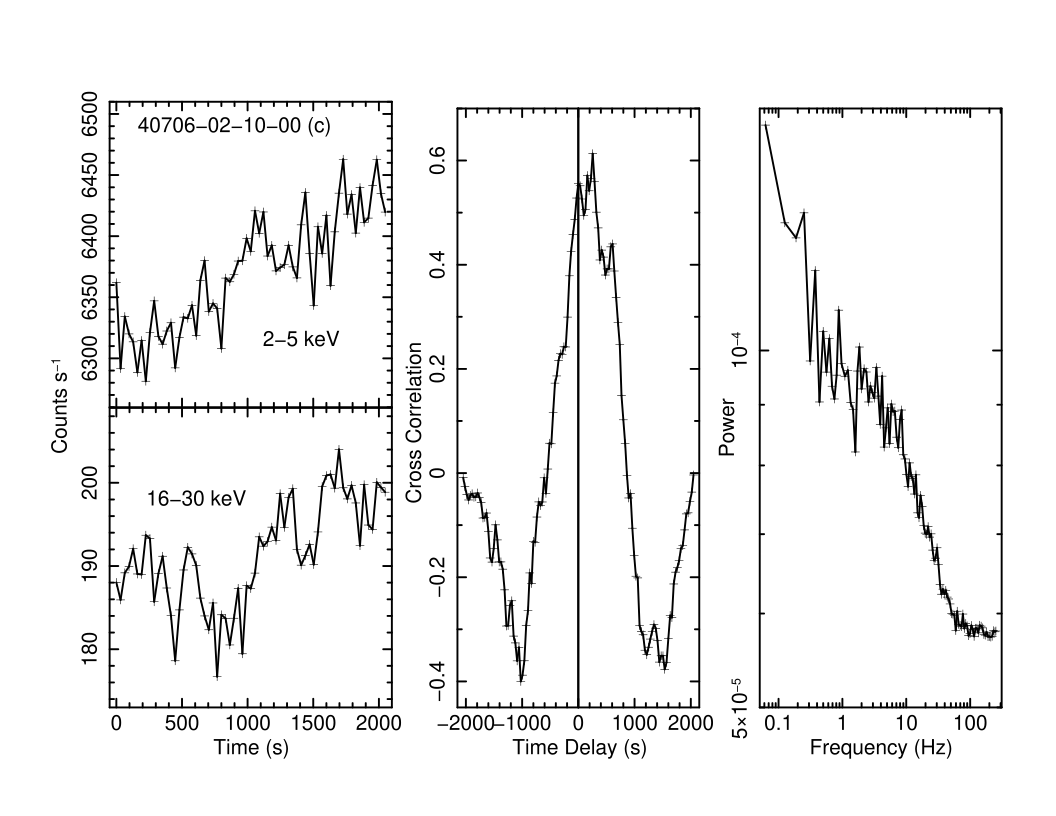} \\ 

\includegraphics[height=7 cm, width=17 cm, angle=0]{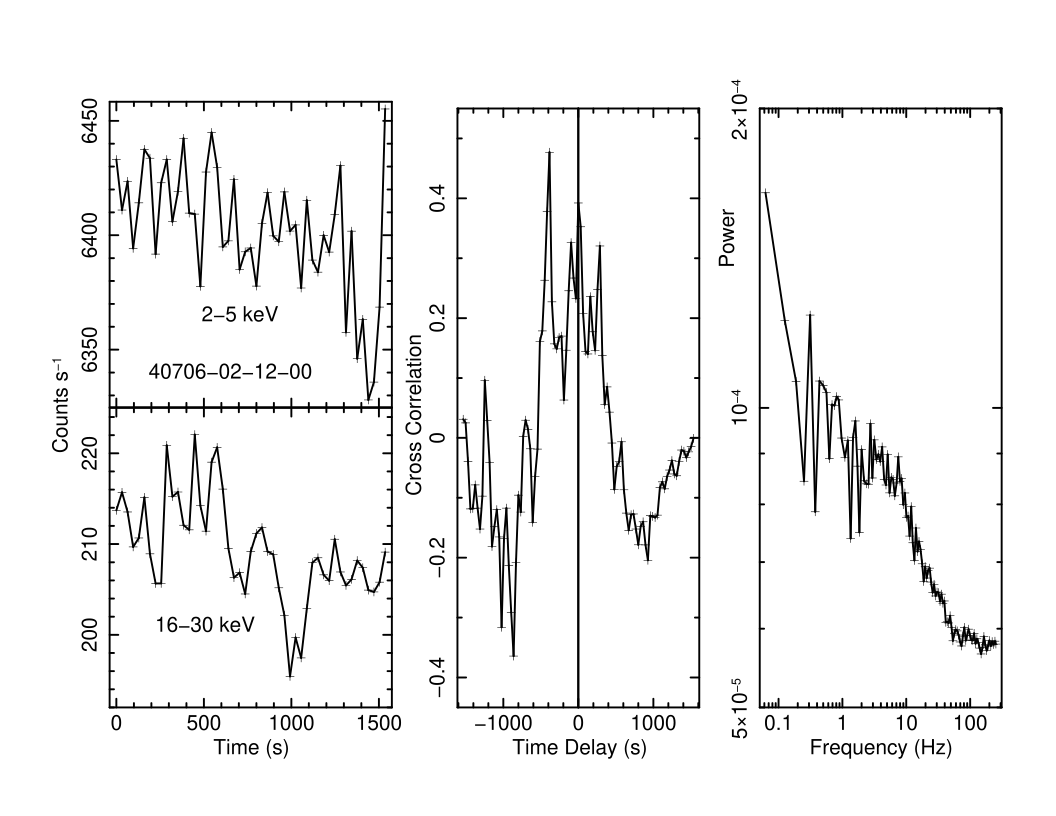} \\

\caption{Each panel displays the soft (2-5 keV) and hard (16-30 keV) energy band light curves (left panels) along with cross-correlation functions (CCF) in the right panels. The vertical line shows the zero lag for all observations.}
\end{figure*}

\clearpage
\begin{figure*}
\includegraphics[height=7 cm, width=17 cm, angle=0]{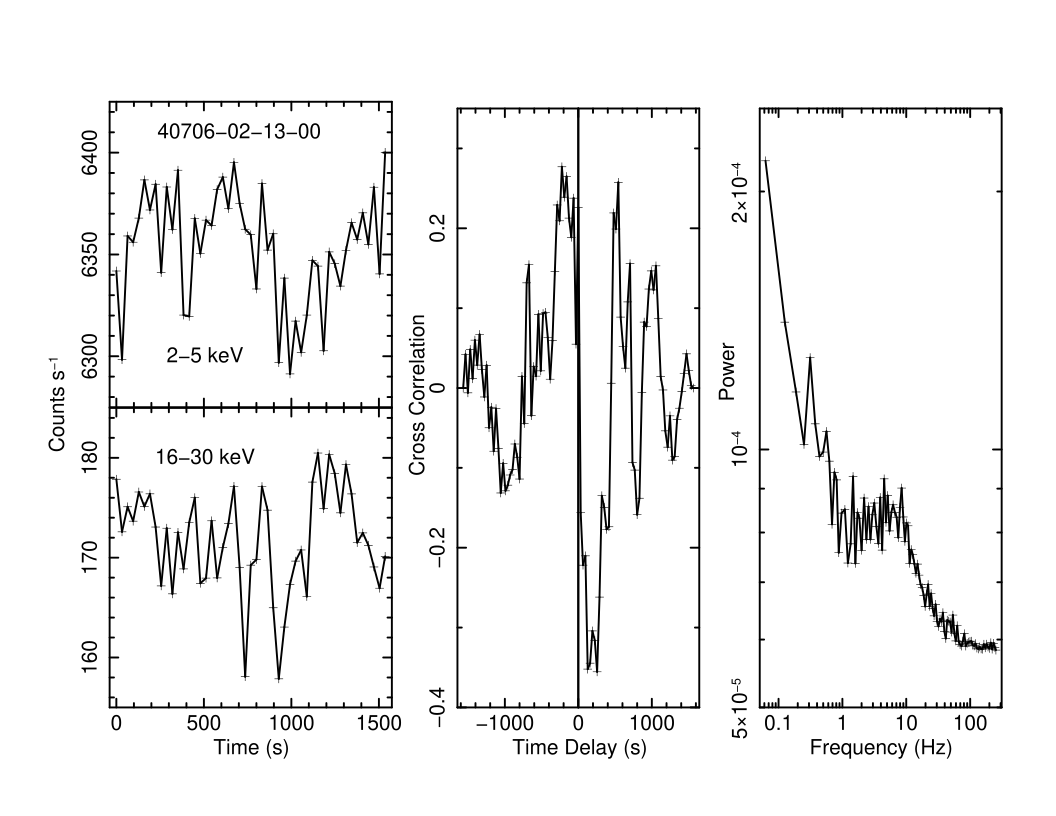} \\ 

\includegraphics[height=7 cm, width=17 cm, angle=0]{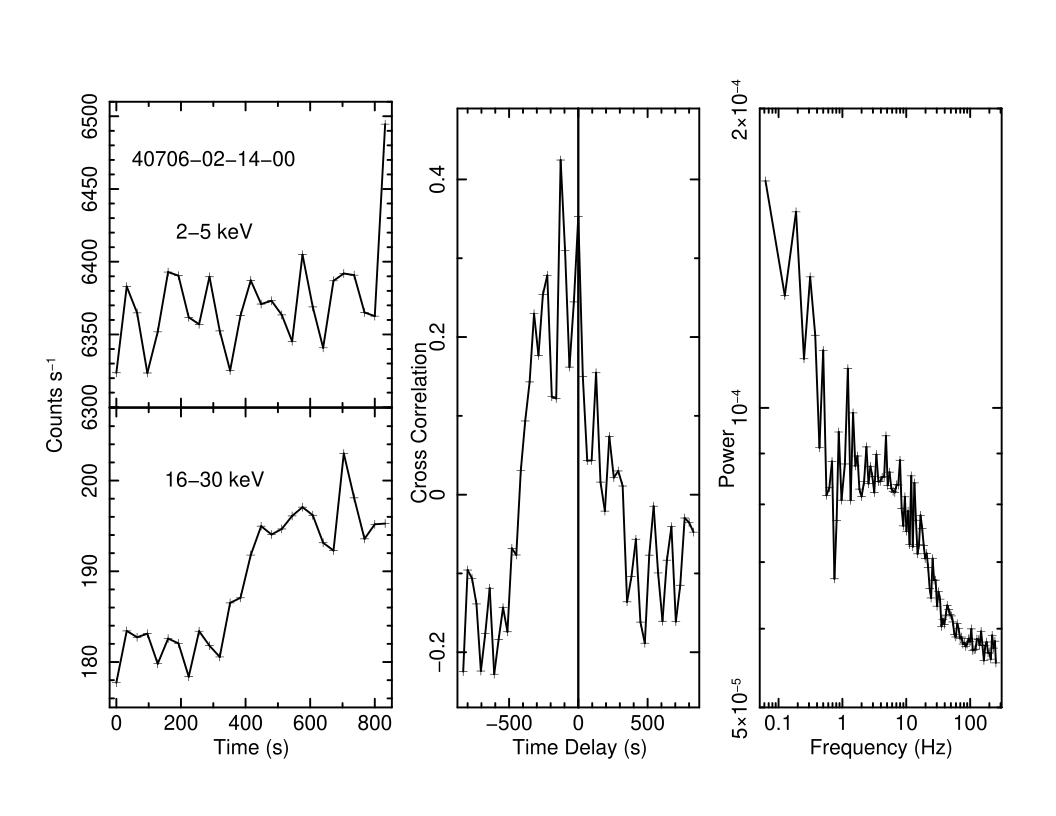} \\ 
{{\bf Fig. 3}:             Continue.. }
\end{figure*}


\clearpage
\begin{figure*}
\includegraphics[height=5 cm, width=9cm, angle=0]{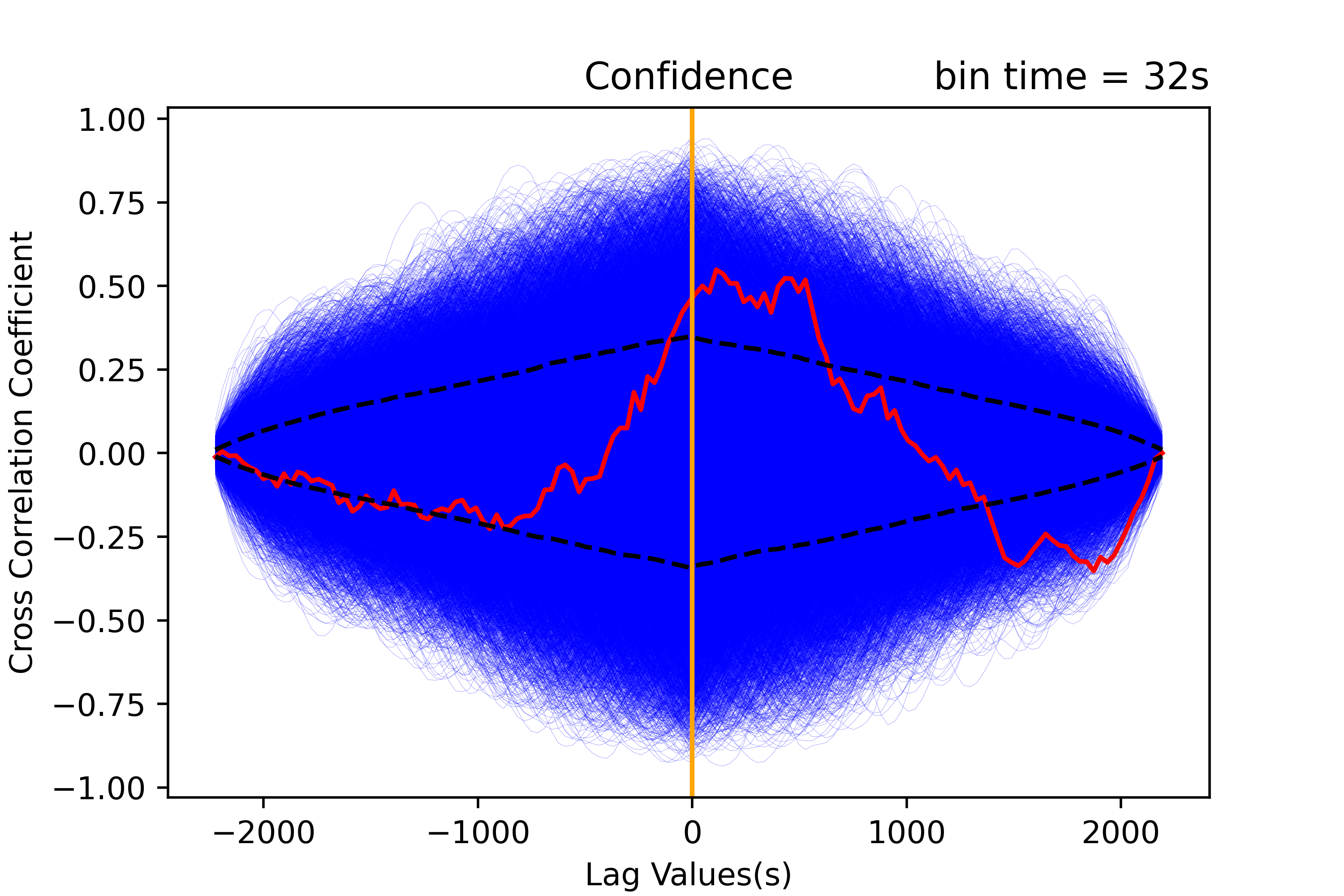} 
\includegraphics[height=5 cm, width=9 cm, angle=0]{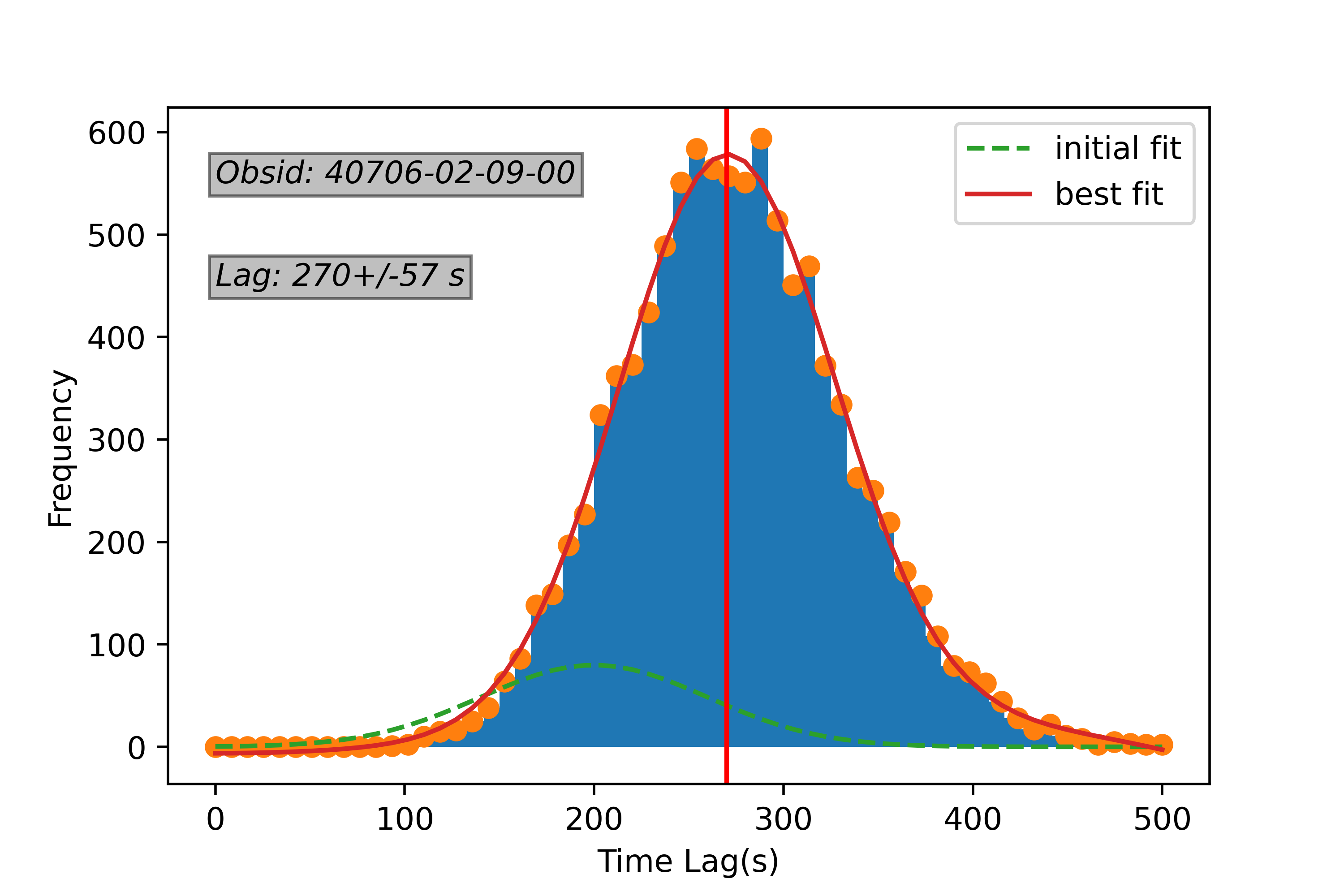}\\

\includegraphics[height=5 cm, width=9cm, angle=0]{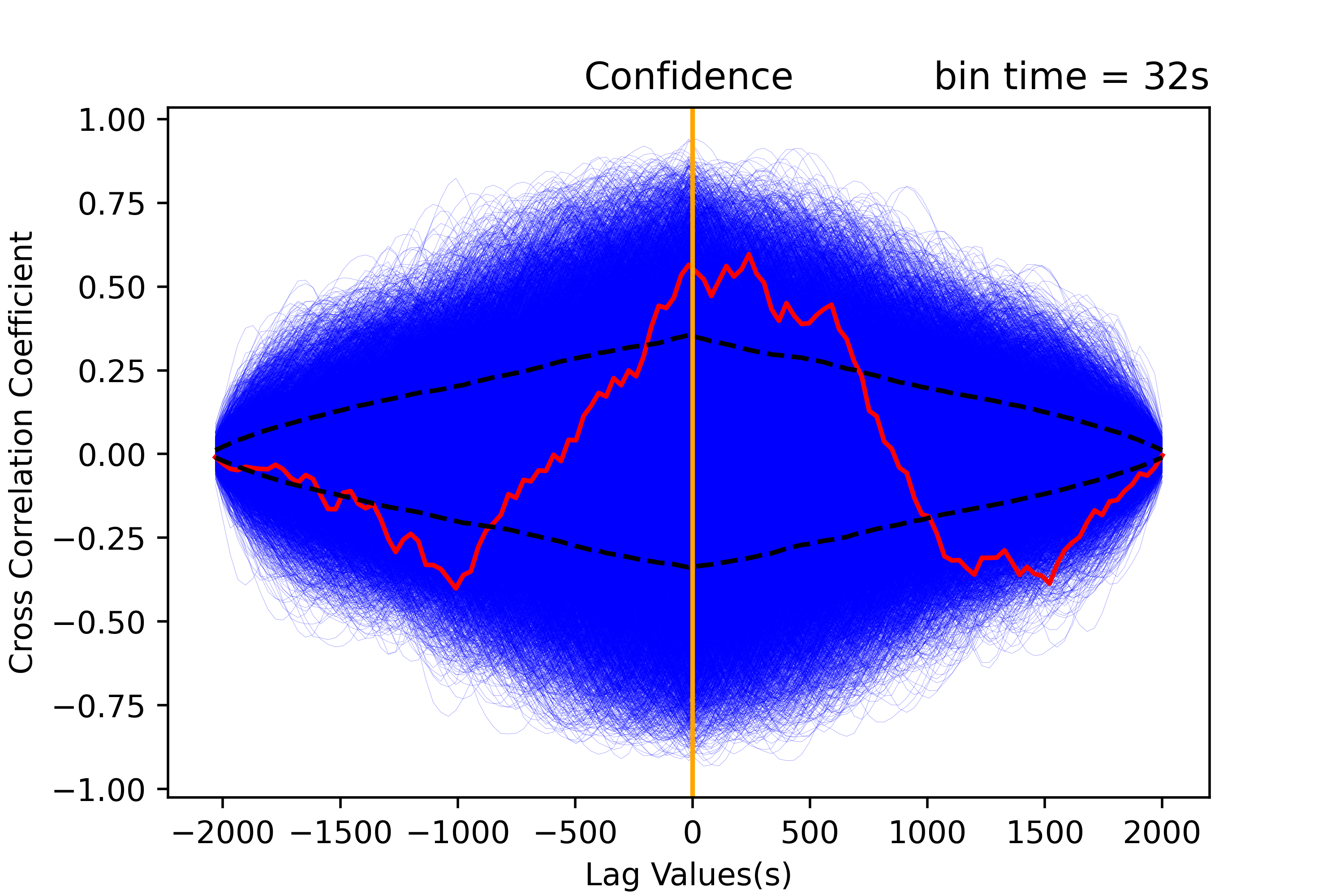} 
\includegraphics[height=5 cm, width=9 cm, angle=0]{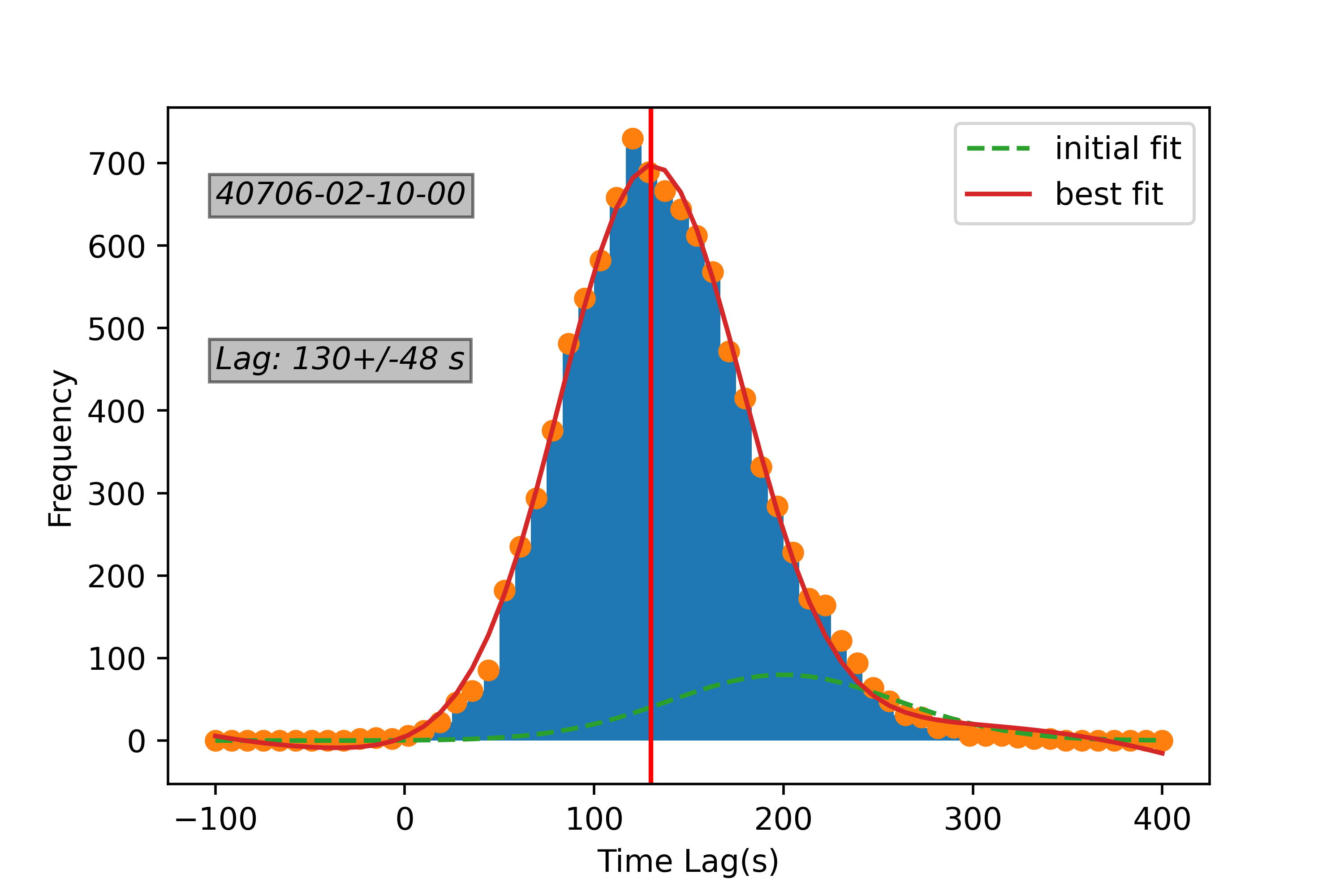}\\

\includegraphics[height=5 cm, width=9cm, angle=0]{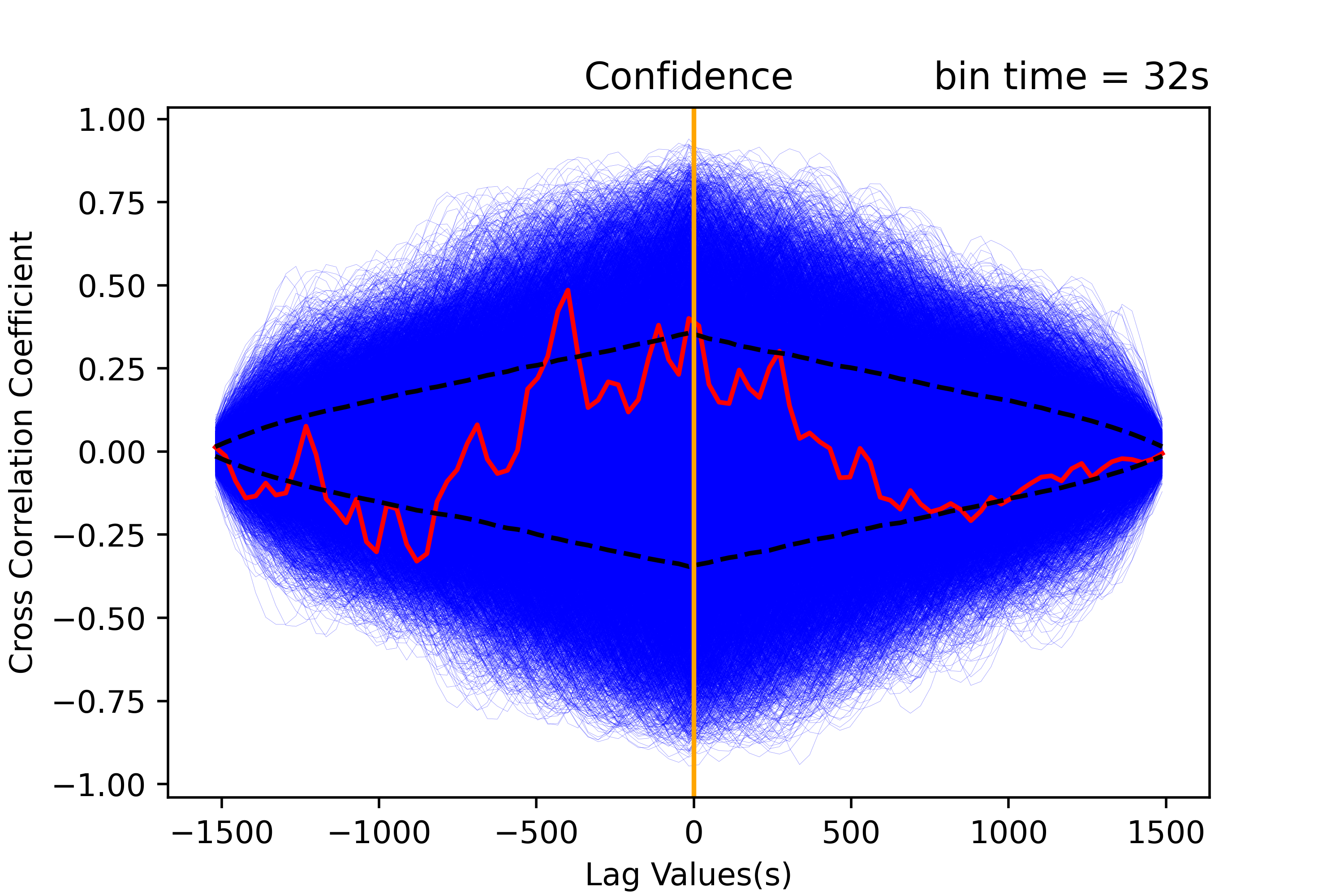} 
\includegraphics[height=5 cm, width=9 cm, angle=0]{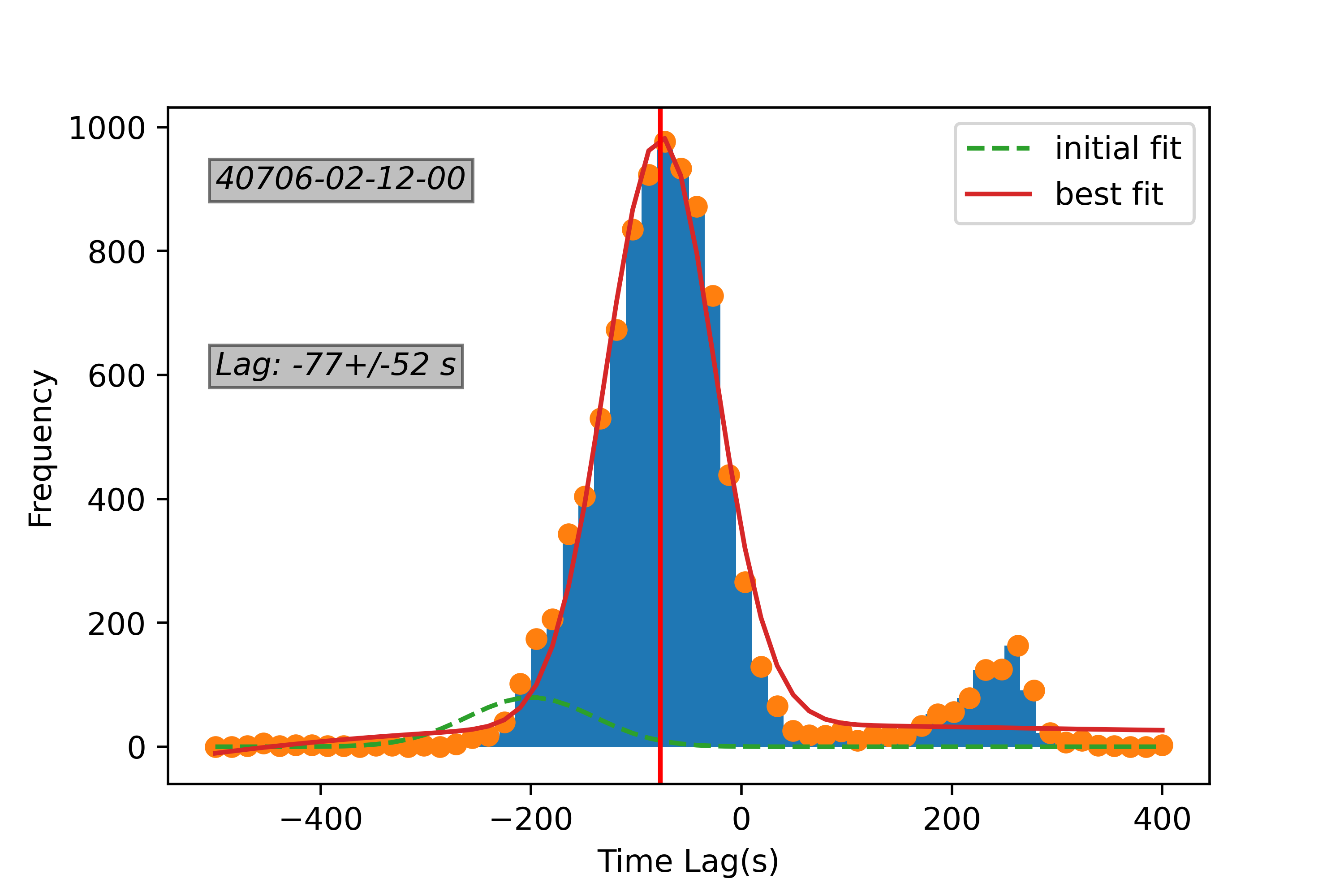}\\

\includegraphics[height=5 cm, width=9cm, angle=0]{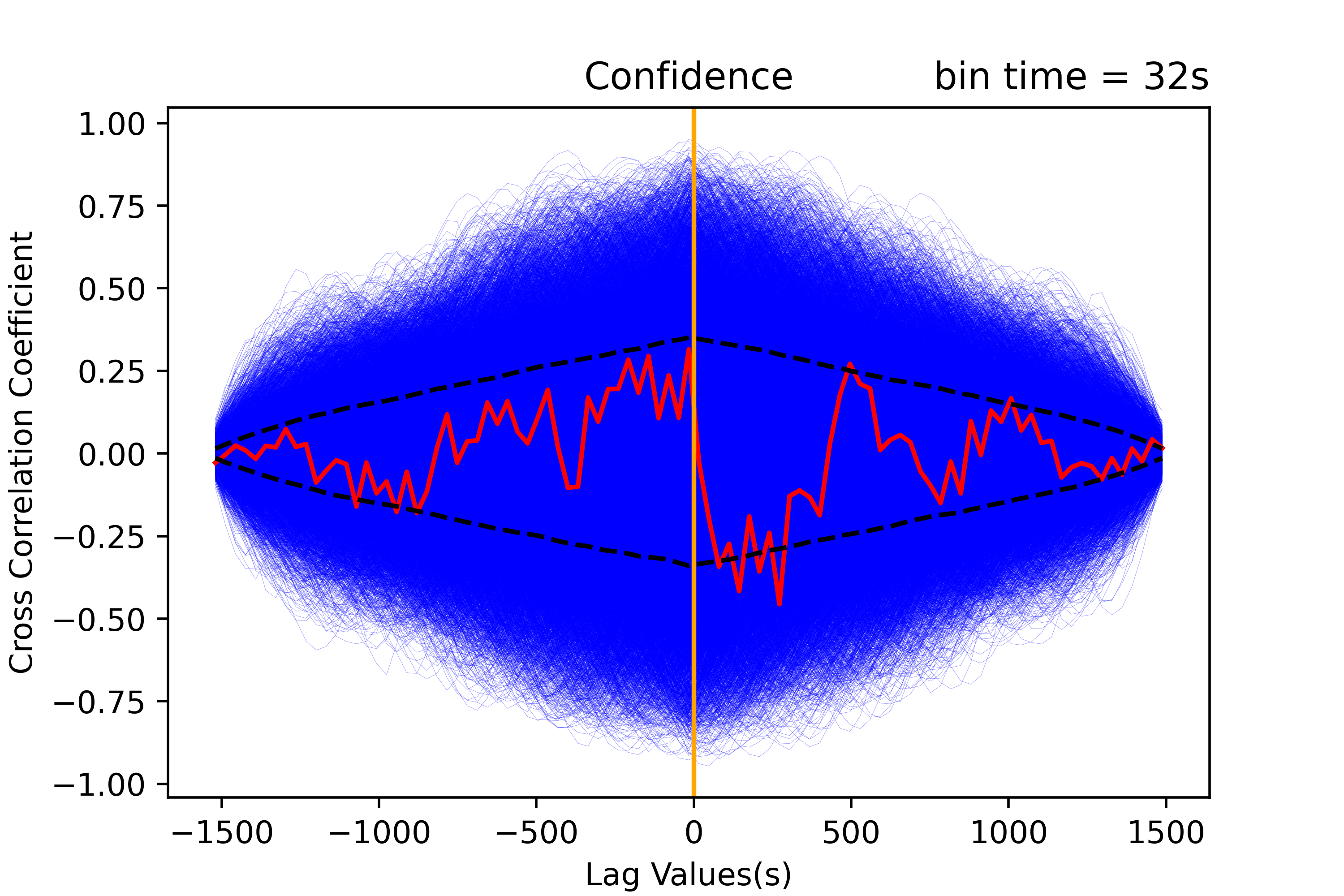} 
\includegraphics[height=5 cm, width=9 cm, angle=0]{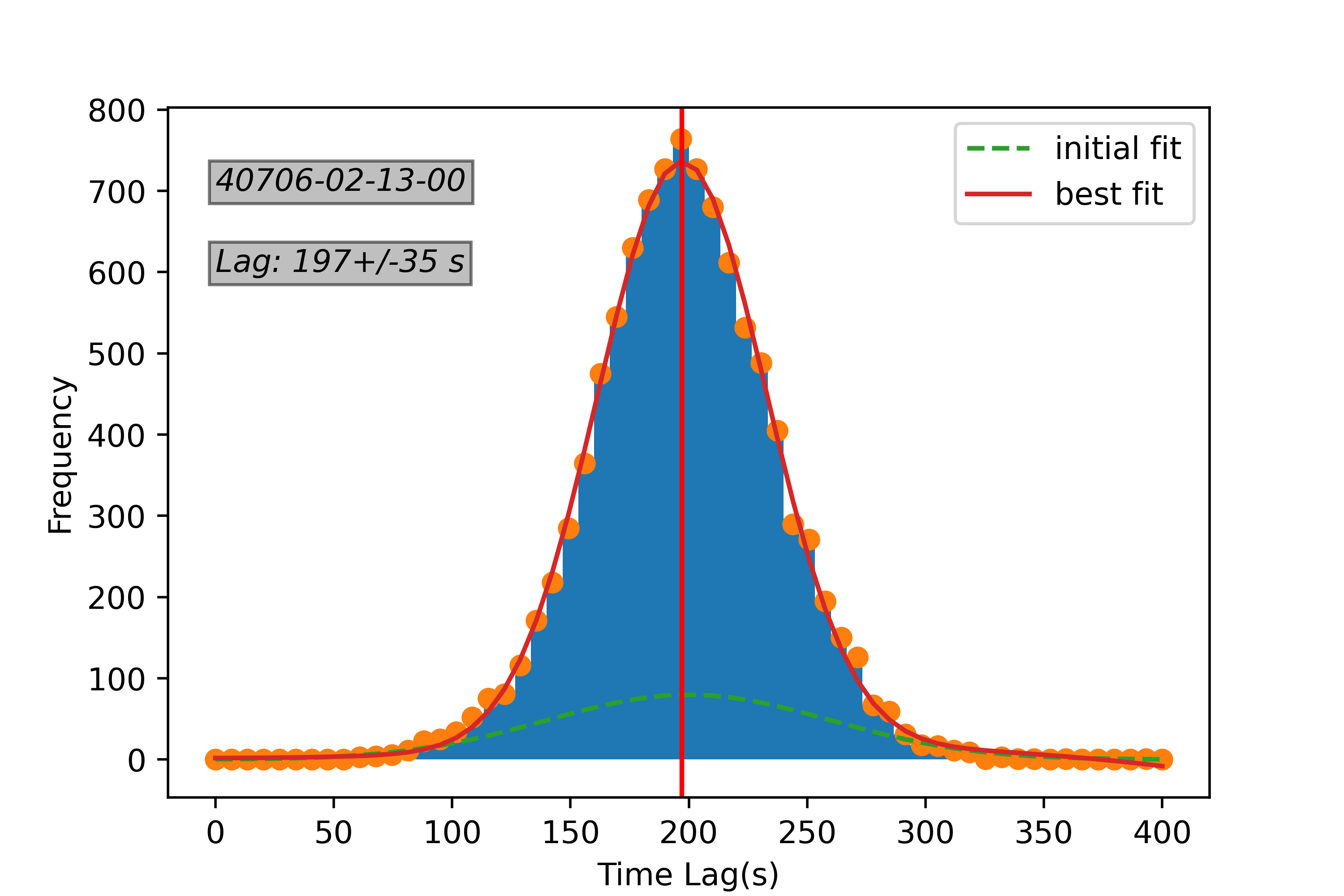}\\

\caption{Left panels blue shaded region displays the 10000 simulated CCFs evaluated from their respective soft (2-5 keV) ad hard (16-30 keV) energy band light curves along with the 95\% significance marked with dashed lines. The observed CCF is shown with a red line along with the vertical line representing the zero lags. The corresponding lag histogram is shown in the right panels (for more details see the text). The green dashed line is the initial parameters fit and later converges to best fit shown in the red lines.}
\end{figure*}

\clearpage

\begin{figure*}
\includegraphics[height=11cm, width=9cm, angle=0]{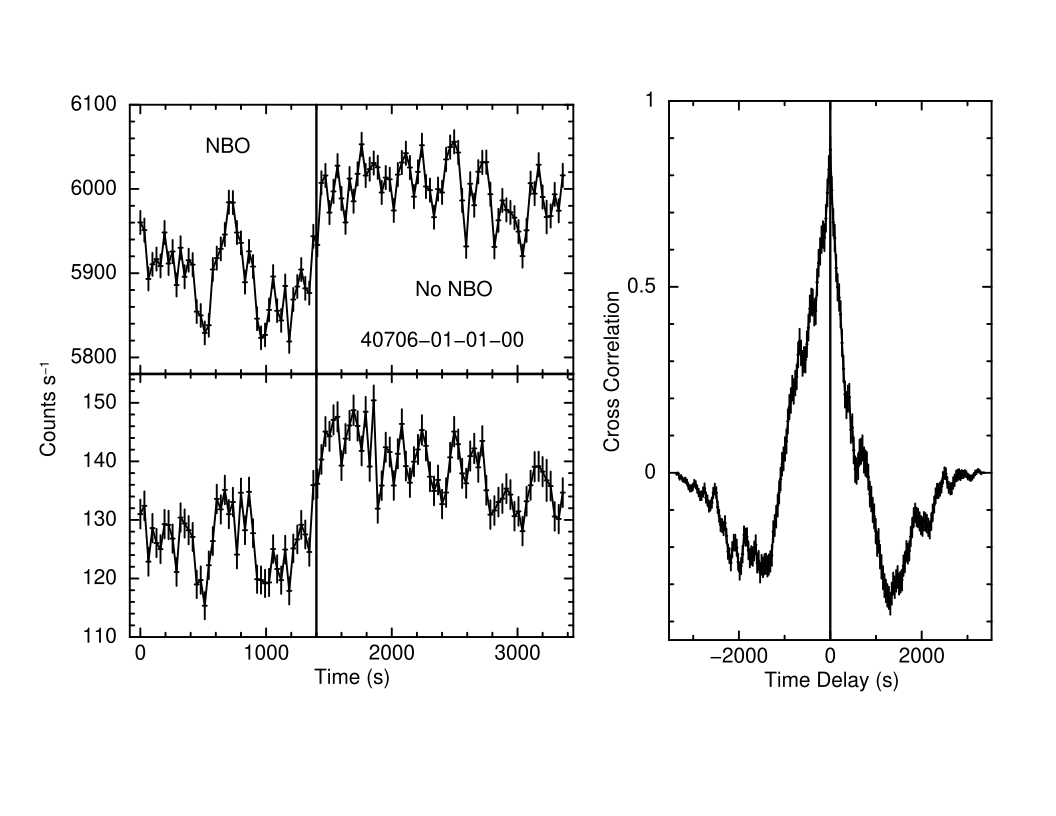} \includegraphics[height=11cm, width=9cm, angle=0]{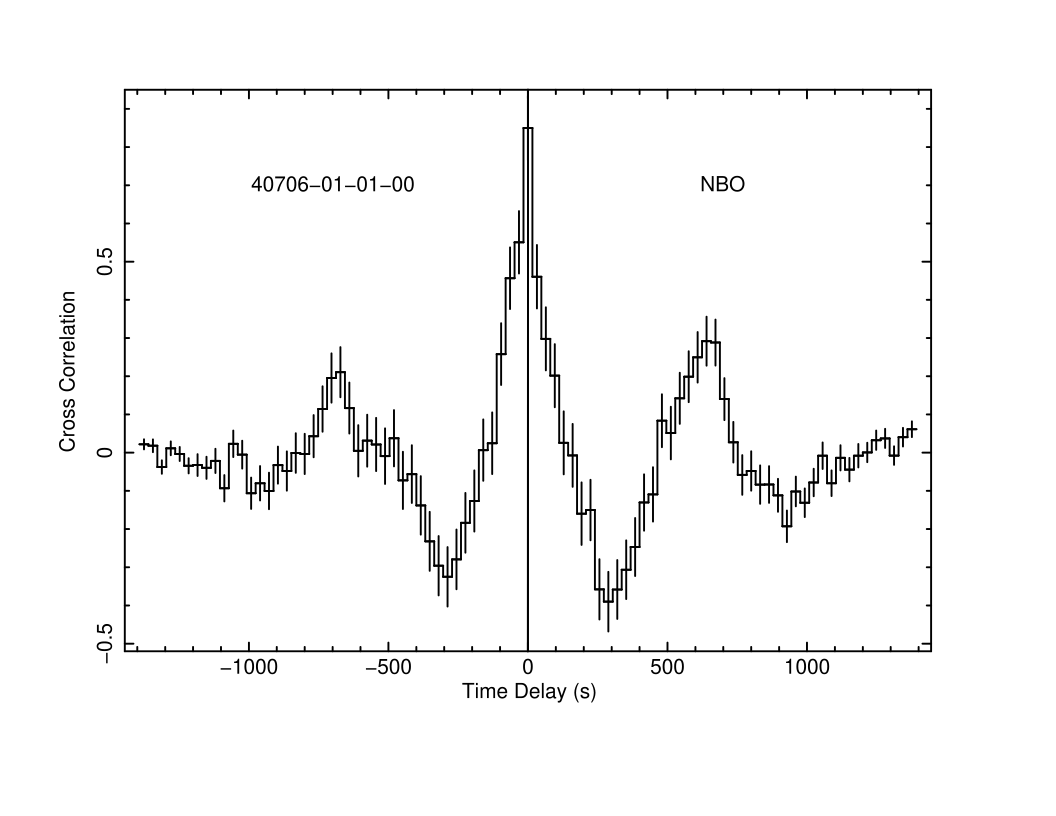}
\includegraphics[height=10cm, width=9cm, angle=0]{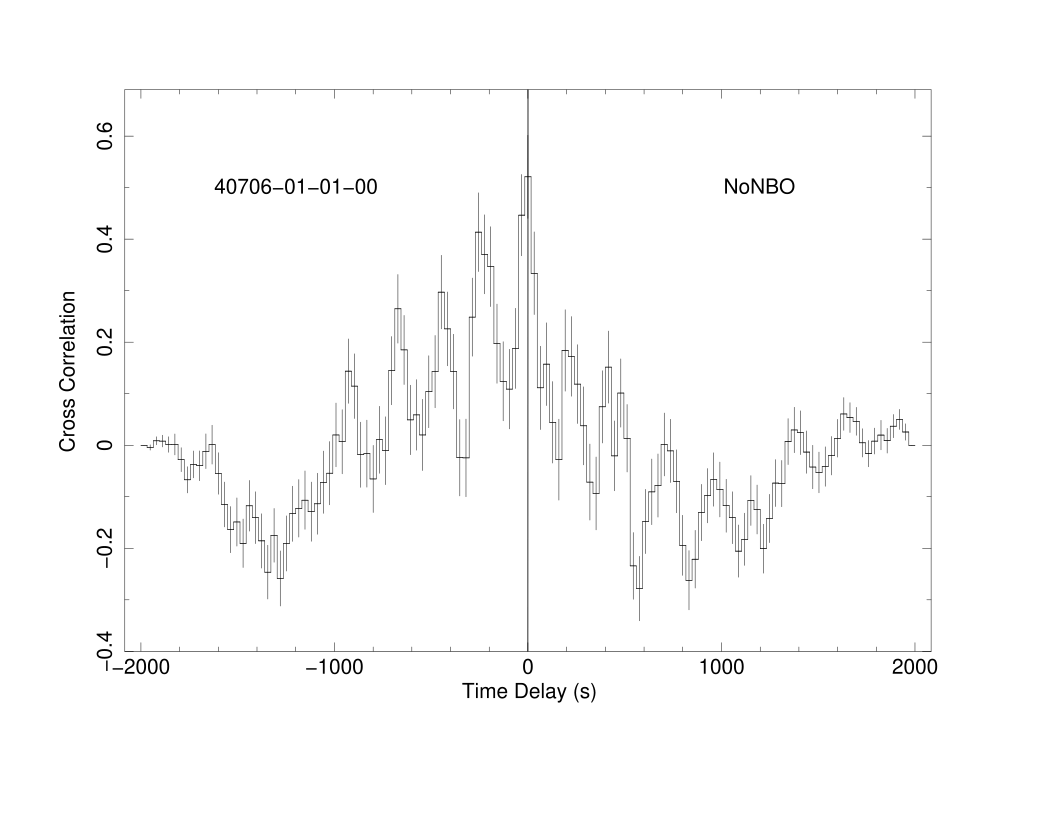} 
\includegraphics[height=10cm, width=9cm, angle=0]{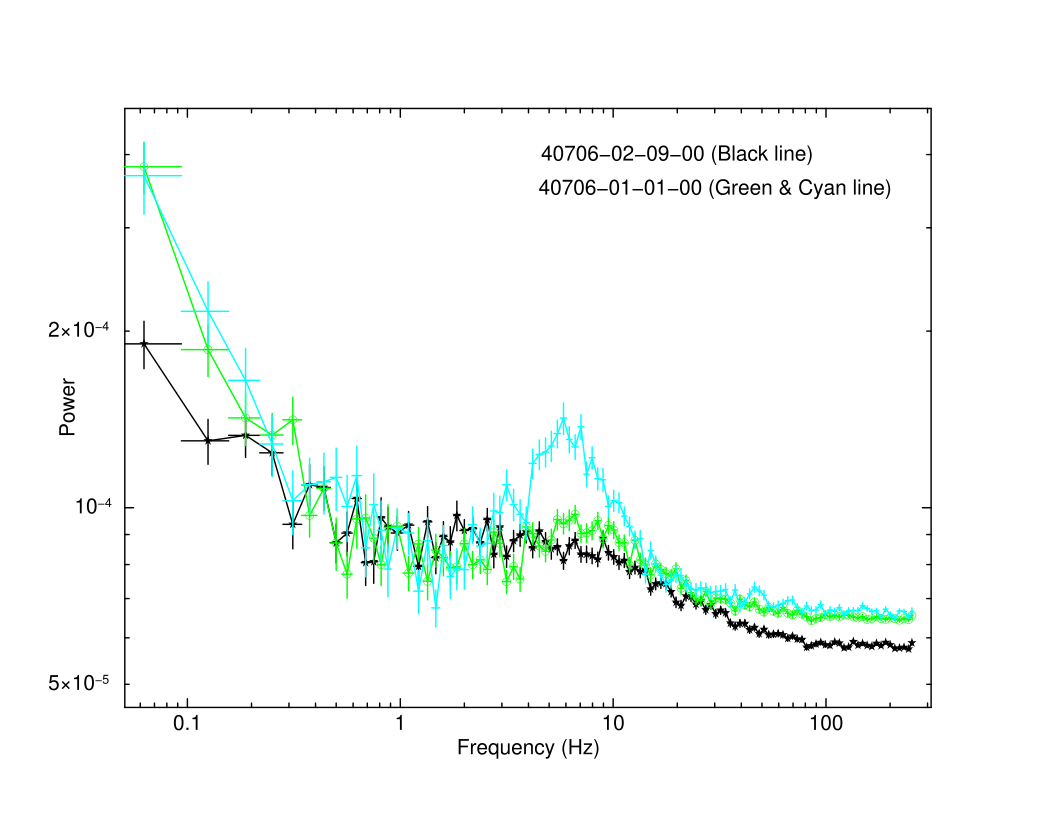}\\\\

\caption{The ObsId. 40706-01-01-00 light curve in 2-5 keV and 16-30 keV, CCF (top-left panel) along with CCFs of NBO (top-right) and No NBO (Bottom-left panel). The bottom right panel shows the PDS of NBO and No NBO sections of the light curve and has been compared with the PDS of ObsId. 40706-02-09-00. This light curve variation occurred before the ObsID. 40702-02-09-00 where lags were detected (Fig. 2, top panel). It is observed that the PDS with No NBO 40706-01-01-00 and 40706-02-09-00 are the same. The bottom left and right panels are the same as above for ObsId. 40702-01-02-00.}

\end{figure*}
\clearpage

\begin{figure*}
\includegraphics[height=10cm, width=9cm, angle=0]{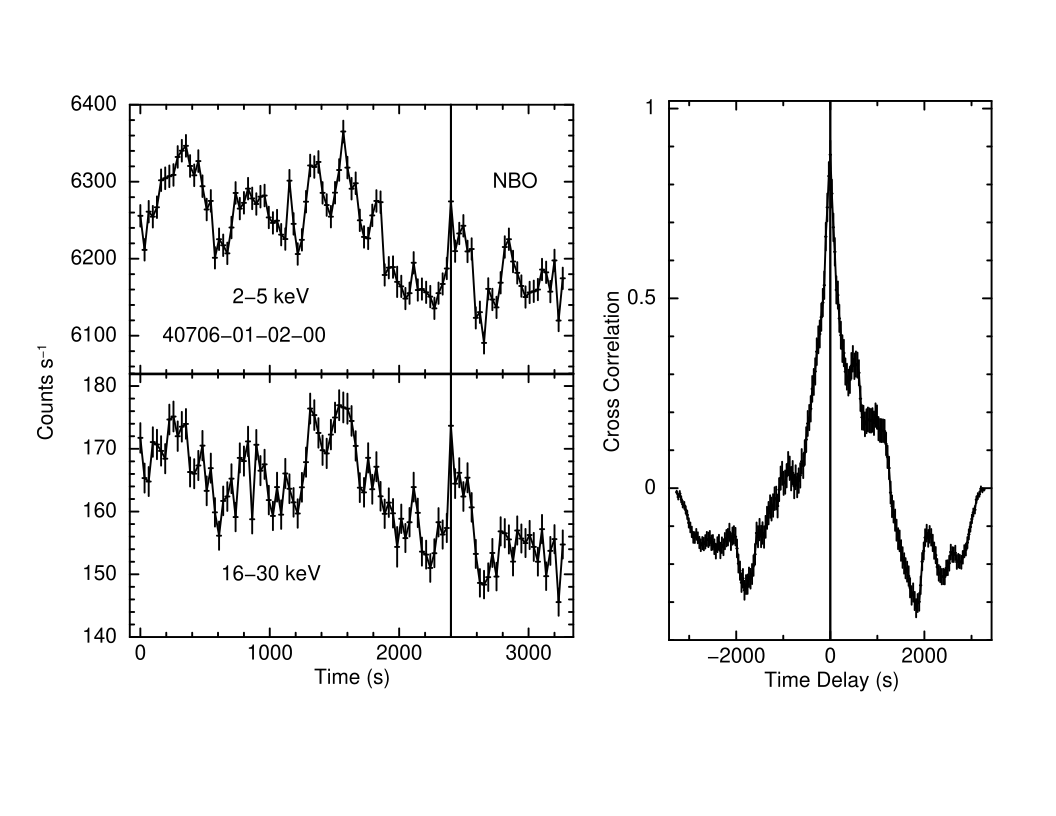} 
\includegraphics[height=10cm, width=9cm, angle=0]{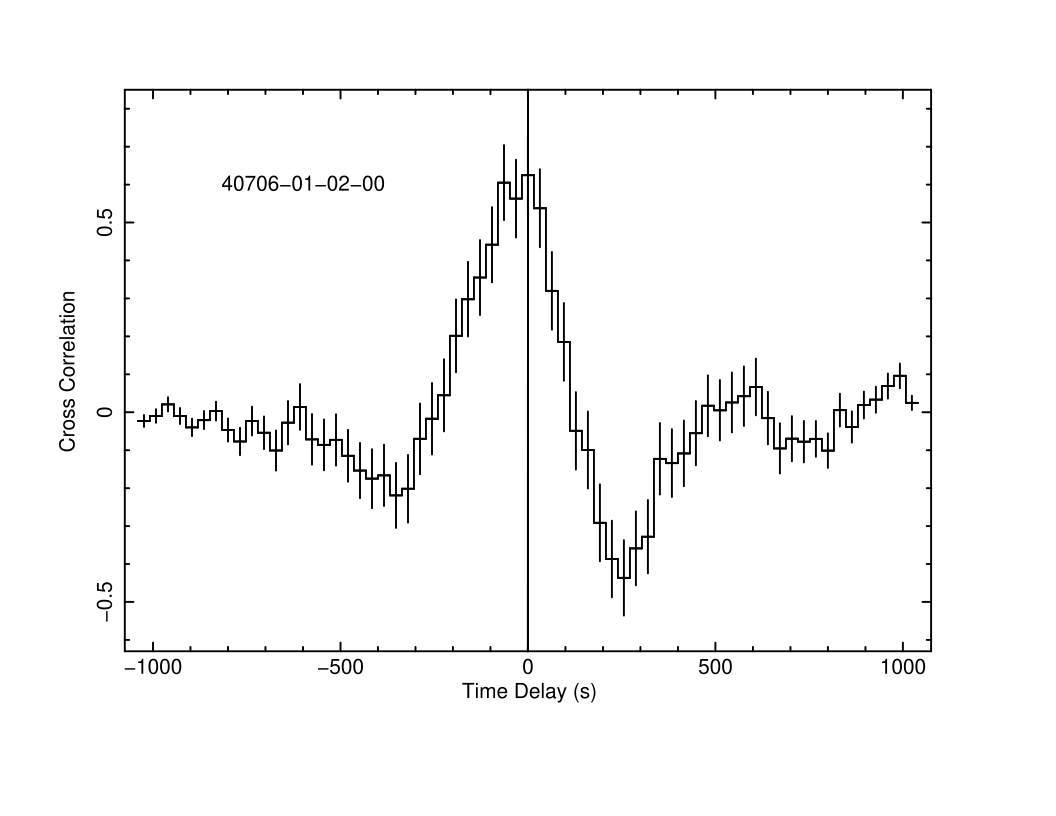} \\ \\
\includegraphics[height=10cm, width=9cm, angle=0]{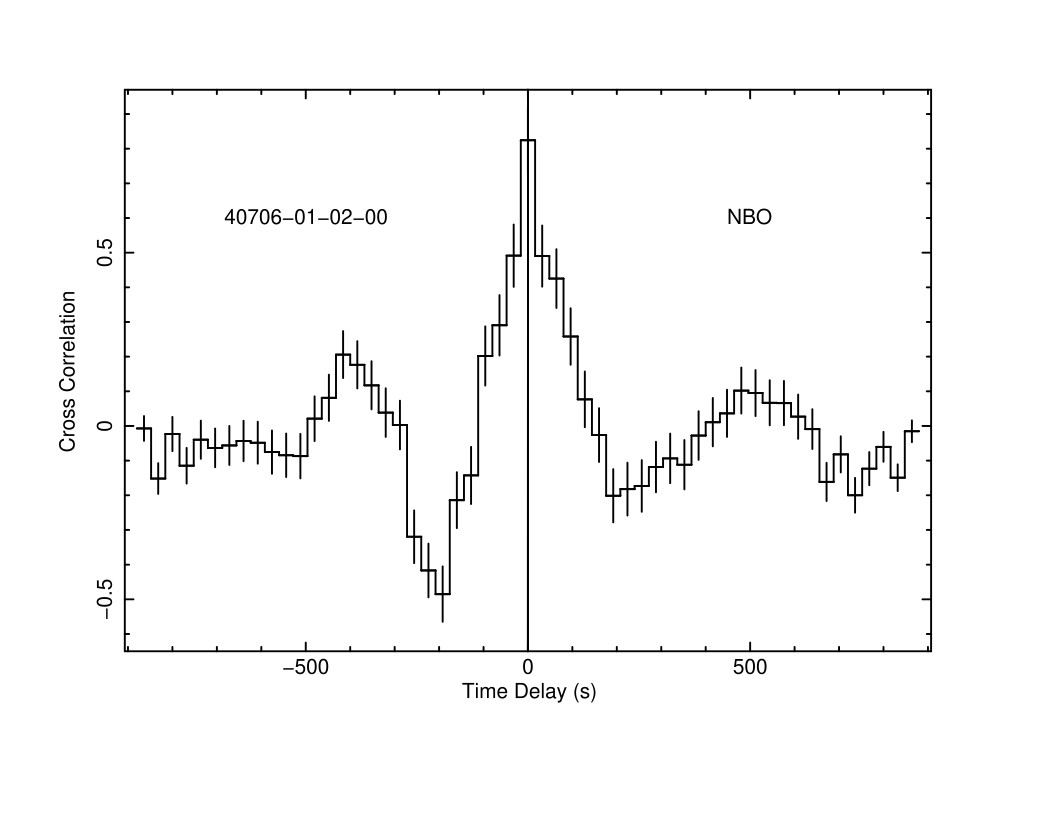} 
\includegraphics[height=10cm, width=9cm, angle=0]{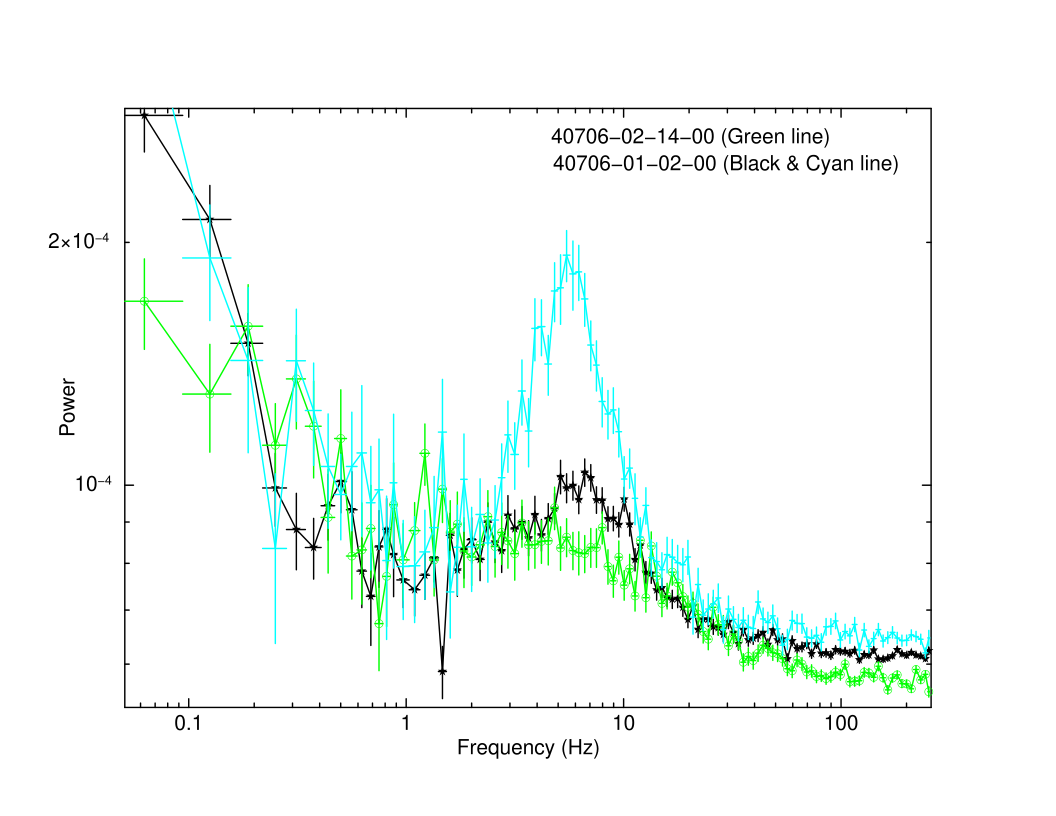}\\\\

\caption{Same as above for the $\dagger$11 (40706-01-02-00). A delay of $\sim$ 90 s is seen in the NoNBO section of the light curve, which is absent in the NBO section. The flat-top noise feature in the PDS (black) is similar to that of the PDS observed in $\dagger$10 (40706-01-14-00, green). }

\end{figure*}
\begin{figure*}
\includegraphics[height=8.5cm, width=10cm, angle=90]{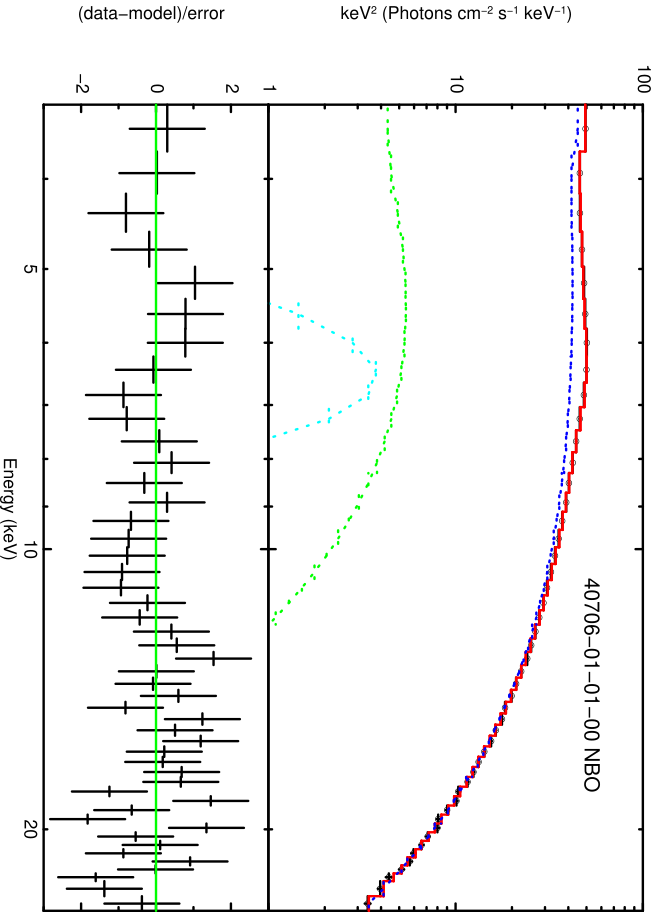}
\includegraphics[height=10cm, width=8.5cm, angle=0]{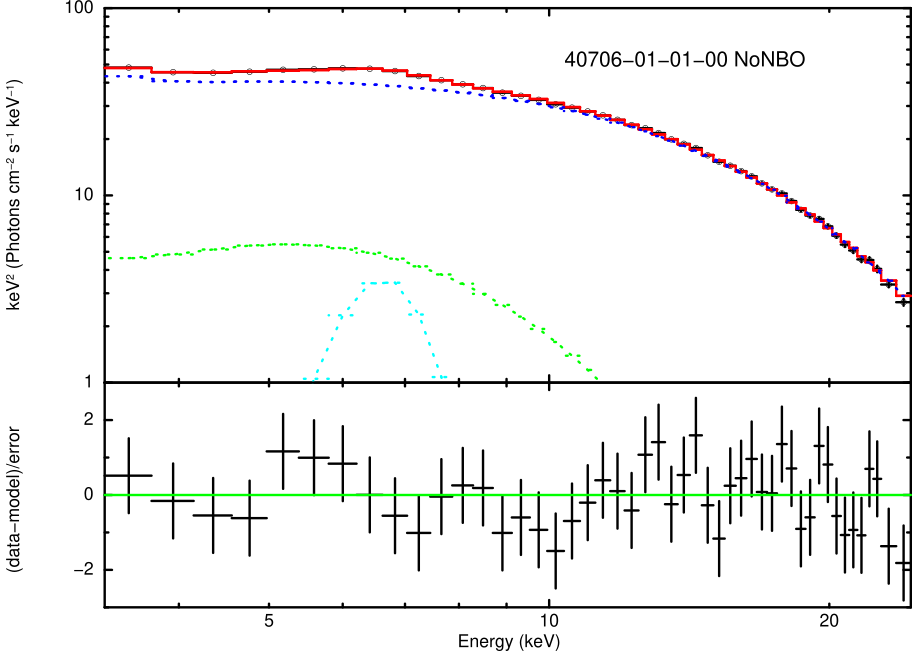}\\
\end{figure*}
\begin{figure*}
\includegraphics[height=10cm, width=8.5cm, angle=0]{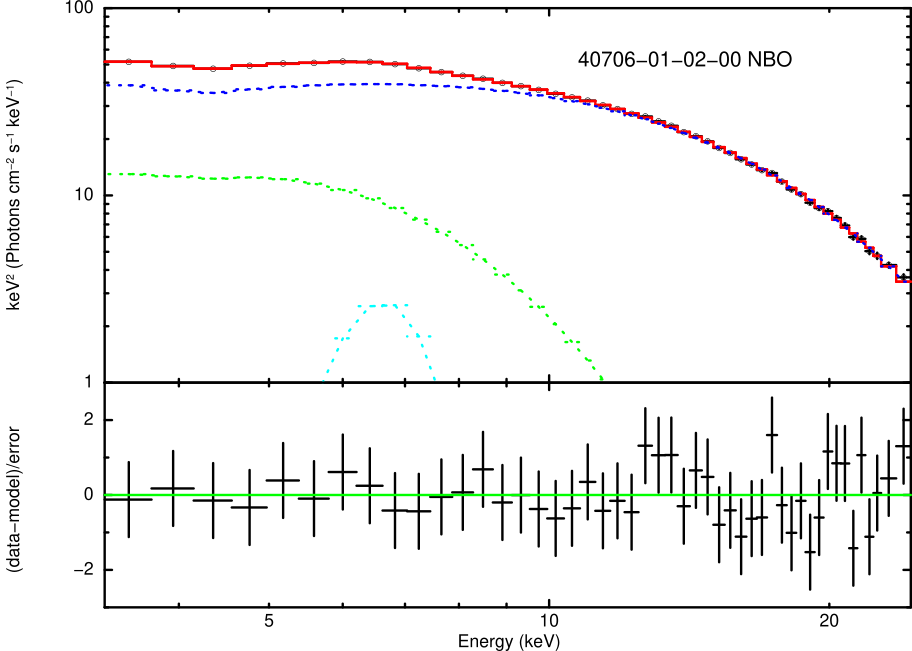}
\includegraphics[height=10cm, width=8.5cm, angle=0]{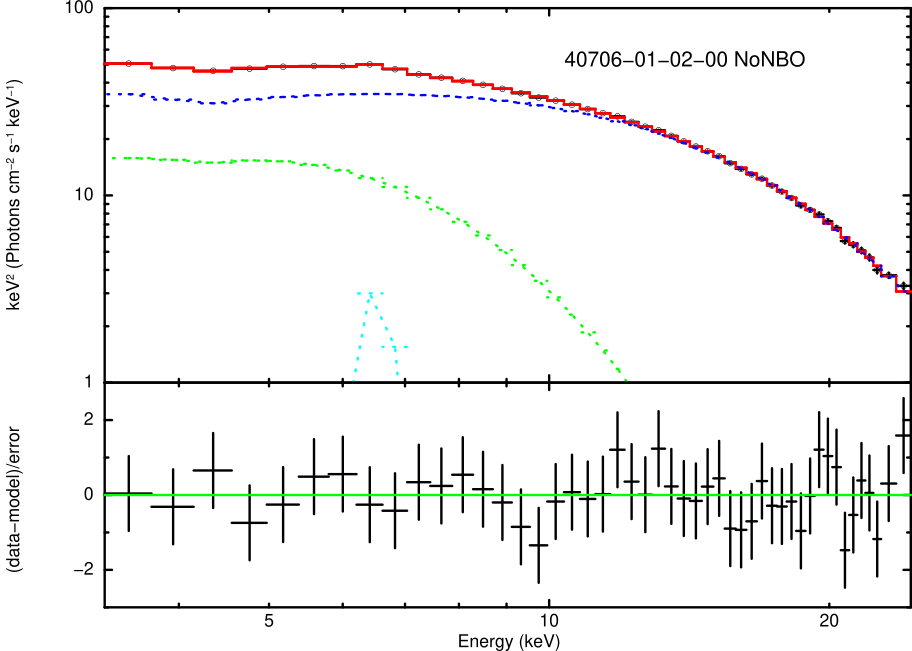}\\\\
\caption{Best-fit spectra for both NBO and NoNBO sections of the respective observations (see the inset in each panel) using the model "bbody+nthcomp+gaussian". Black points are observations, the red line is the continuum model fit, the blue line is nthcomp model component, the green line is the black body component and the cyan is the Gaussian for Fe spectral line. The lower sub-panels display the residuals. }

\end{figure*}

\clearpage
\newpage

\begin{table}[!h]
    \centering
    \begin{adjustbox}{width=1\textwidth}
    \renewcommand{\arraystretch}{2}
    \begin{tabular}{|c|>{\centering\arraybackslash}p{0.3\linewidth}|>{\centering\arraybackslash}p{0.25\linewidth}|c|c|>{\centering\arraybackslash}p{0.3\linewidth}|>{\centering\arraybackslash}p{0.3\linewidth}|>{\centering\arraybackslash}p{0.3\linewidth}|c|c|c|} \hline 
    Sr. No $\dagger$&  MJD
(Motta \& Fender 2019)&Radio Event
(Fomalont 2001 a, b)&X-ray ObsId. &Branch &NBO frequency 
(Hz)&  HBO frequency
(Hz)&FBO frequency 
(Hz) &Flat topped
noise& Delay (s)
&CC\\ \hline 

 &  51339.254$\pm$0.003&Lobe Ejection 1& &  & & &  && &\\ 
          1&  51339.384&&40706-02-01-00 & NB&$6.03^{+0.03}_{-0.04}$&  $44.6^{+0.64}_{-0.72}$ &-  &-& - & 0.92$\pm$0.05\\  
          2&  51339.645&&40706-02-03-00 & NB&$6.32^{+1.12}_{-1.14}$&  - &-  &-& -& 0.88$\pm$0.08\\ 
          3&  51339.977&&40706-02-06-00 & NB&$9.14^{+0.08}_{-0.07}$&  - &$16.74^{+0.20}_{-0.23}$  &-& -&0.82$\pm$0.07\\
 &  51340.39$\pm$0.02&Core Flare 1& &  & & &  && &\\  
          4&  51340.425&&40706-01-01-00 & NB&$8.77^{+0.61}_{-0.53}$&  - &-  &-& -&0.90$\pm$0.10\\
 &  51340.779$\pm$0.004&Lobe Ejection 2& &  & & &  && &\\  

          5&  51340.839&&40706-02-09-00 & HB&-&  - &-  &Yes&$270\pm{57}$
&$0.55\pm{0.06}$\\
 &  51340.86$\pm$0.02&Core Flare 2& &  & & &  && &\\ 
          6&  51340.909&&40706-02-08-00 & HB/Hard Apex&-&  - &-  &Yes& - & $-0.18\pm0.03$\\  
          7&  51340.975&&40706-02-10-00 & HB/Hard Apex&-&  - &-  &Yes& $130\pm{48}$
&$0.51\pm{0.06}$\\ 
          8&  51341.159&&40706-02-12-00 & HB/Hard Apex&-&  - &-  &Yes& $-77\pm{52}$
&$0.48\pm{0.17}$\\  
          9&  51341.225&&40706-02-13-00 & HB/Hard Apex&-&  - &-  &Yes& $197\pm{35}$
&$-0.35\pm0.14$\\  
        10&  51341.291&&40706-02-14-00 & HB/Hard Apex&-&  - &-  &Yes& $110\pm{65}$&$-0.40\pm0.12$\\
 &  51341.32$\pm$0.02&Core Flare 3& &  & & &  && &\\
 &  &Ejection of URF 1& &  & & &  && &\\ 
          11&  51341.358&&40706-01-02-00 & NB&$5.84^{+0.02}_{-0.03}$&  $44.72^{+0.91}_{-0.84}$ &-  &-& -&$-0.85\pm0.10$\\
 &  51341.42$\pm$0.02&Lobe Flare N3& &  & & &  && &\\ 
          12&  51341.577&&40706-02-16-00 & NB&-&  - &$17.27^{+0.26}_{-0.23}$  &-& -&$-0.95\pm0.07$\\  
          13&  51341.699&&40706-02-17-00 & NB&$8.91^{+0.14}_{-0.28}$&  - &$12.27^{+0.21}_{-0.33}$  &-& -&$-0.38\pm0.05$\\  
          14&  51341.906&&40706-02-18-00 & NB&$8.14^{+1.01}_{-0.97}$&  - &-  &-& -&$-0.80\pm0.06$\\ 
  15&  51341.973&&40706-02-19-00 & NB&$6.18^{+0.08}_{-0.10}$& $43.12^{+0.69}_{-0.72}$ &-  &-& -&$-0.92\pm0.07$\\
 &  51341.99$\pm$0.02&Core Flare 4& &  & & &  && &\\
 &  &Ejection of URF 2& &  & & &  && &\\
 &  51342.13$\pm$0.02&Lobe Flare S3& &  & & &  && &\\  
          16&  51342.158&&40706-02-21-00 & NB&-&  - &-  &Yes& -&$-0.95\pm0.06$\\  
          17&  51342.291&Low Core Radio Flux&40706-02-23-00 & NB&$9.21^{+0.42}_{-0.31}$&  - &$17.67^{+0.45}_{-0.41}$  &-& -&$-0.81\pm0.04$\\

 &  51342.3$\pm$0.02&Lobe Flare N4& &  & & &  && &\\\hline
    \end{tabular}
    \end{adjustbox}
    \begin{minipage}{\textwidth}
    \caption{Log of observations where Sco X-1 was simultaneously observed in X-ray and radio bands (see Motta \& Fender 2019). The radio events are taken from the works of (Fomalont 2001 a, b), where the lobe flares N and S represent the North East and South West lobes. The NBO, HBO, and FBO frequencies columns display the centroid frequencies with 90\% error bars. The last two columns display the observed lag in CCF obtained from the simulations and cross-correlation coefficients (CC).} 
    \end{minipage}

\end{table}

\begin{table*}
    \centering
    \begin{adjustbox}{width=1\textwidth}
    \begin{tabular}{|l|ll|ll|}
    \hline
    ObsID&\hspace{0.2cm}40706-01-01-00 ($\dagger$ 4)&&\hspace{0.2cm}40706-01-02-00 ($\dagger$ 11)&\\
    \hline
    Parameters&NBO&NoNBO&NoNBO&NBO\\
    
    \hline
    kT$_{bb}$ (keV)&1.35$\pm$0.03&1.30$\pm$0.04&1.42$\pm$0.04&1.41$\pm$0.03\\
    N$_{bb}$&0.14$\pm$0.01&0.15$\pm$0.01&0.14$\pm$0.01&0.17$\pm$0.01\\
    $\Gamma$$_{nthcomp}$&1.94$\pm$0.01&2.01$\pm$0.01&1.97$\pm$0.01&2.02$\pm$0.01\\
    kT$_{e}$ (keV)&2.87$\pm$0.02&2.82$\pm$0.02&2.90$\pm$0.02&2.94$\pm$0.02\\
    N$_{nthcomp}$&25.04$\pm$0.52&29.52$\pm$0.57&27.47$\pm$0.56&27.85$\pm$0.59\\
    kT$_{soft}$ (keV)&0.32 $\pm$ 0.02 &0.36 $\pm$ 0.02 &0.32$\pm$ 0.02&0.35$\pm$ 0.02\\
    Flux&10.96$\pm$0.01&11.45$\pm$0.01&11.88$\pm$0.01&11.22$\pm$0.01\\
    Flux$_{bb}$&1.17$\pm$0.02&1.29$\pm$0.02&3.17$\pm$0.02&2.52$\pm$0.02\\
    Flux$_{nthcomp}$&9.79$\pm$0.01&10.16$\pm$0.01&8.71$\pm$0.01&8.70$\pm$0.01\\
    {\it($\Delta$f/f)$_{bb}$}&\hspace{2cm}0.102&&\hspace{2cm}0.205&\\
    {\it($\Delta$f/f)$_{ntcomp}$}&\hspace{2cm}0.037&&\hspace{2cm}0.001&\\

    $\chi^{2}$/dof&37/41&39/41&36/41&36/41\\
    \hline
    \end{tabular}
    \end{adjustbox}
    \begin{minipage}{\textwidth}
    \caption{ Best fit spectral parameters using  bbody+gaussian+nthcomp in 3-25 keV band. kT$_{bb}$ is the black body temperature along with normalization N$_{bb}$. $\Gamma_{Nthcomp}$ is the asymptotic power-law index and kT$_{e}$ is the electron temperature and kT$_{soft}$ is the soft seed photons arriving from the black body emission of nthcomp model. Fluxes are in the units 10$^{-8}$ erg cm$^{-2}$ s$^{-1}$. ($\Delta$f/f) is the fractional flux variation for the respective model.}
    \end{minipage}
\end{table*}


\cite{*}
\bibliographystyle{spr-mp-nameyear-cnd}
\bibliography{Sco X-1}

\begin{thebibliography}{51}
\ifx \bisbn   \undefined \def \bisbn  #1{ISBN #1}\fi
\ifx \binits  \undefined \def \binits#1{#1}\fi
\ifx \bauthor  \undefined \def \bauthor#1{#1}\fi
\ifx \batitle  \undefined \def \batitle#1{#1}\fi
\ifx \bjtitle  \undefined \def \bjtitle#1{#1}\fi
\ifx \bvolume  \undefined \def \bvolume#1{\textbf{#1}}\fi
\ifx \byear  \undefined \def \byear#1{#1}\fi
\ifx \bissue  \undefined \def \bissue#1{#1}\fi
\ifx \bfpage  \undefined \def \bfpage#1{#1}\fi
\ifx \blpage  \undefined \def \blpage #1{#1}\fi
\ifx \burl  \undefined \def \burl#1{\textsf{#1}}\fi
\ifx \doiurl  \undefined \def \doiurl#1{\textsf{#1}}\fi
\ifx \betal  \undefined \def \betal{\textit{et al.}}\fi
\ifx \binstitute  \undefined \def \binstitute#1{#1}\fi
\ifx \binstitutionaled  \undefined \def \binstitutionaled#1{#1}\fi
\ifx \bctitle  \undefined \def \bctitle#1{#1}\fi
\ifx \beditor  \undefined \def \beditor#1{#1}\fi
\ifx \bpublisher  \undefined \def \bpublisher#1{#1}\fi
\ifx \bbtitle  \undefined \def \bbtitle#1{#1}\fi
\ifx \bedition  \undefined \def \bedition#1{#1}\fi
\ifx \bseriesno  \undefined \def \bseriesno#1{#1}\fi
\ifx \blocation  \undefined \def \blocation#1{#1}\fi
\ifx \bsertitle  \undefined \def \bsertitle#1{#1}\fi
\ifx \bsnm \undefined \def \bsnm#1{#1}\fi
\ifx \bsuffix \undefined \def \bsuffix#1{#1}\fi
\ifx \bparticle \undefined \def \bparticle#1{#1}\fi
\ifx \barticle \undefined \def \barticle#1{#1}\fi
\ifx \bconfdate \undefined \def \bconfdate #1{#1} \fi
\ifx \botherref \undefined \def \botherref #1{#1} \fi
\ifx \url \undefined \def \url#1{\textsf{#1}} \fi
\ifx \bchapter \undefined \def \bchapter#1{#1} \fi
\ifx \bbook \undefined \def \bbook#1{#1} \fi
\ifx \bcomment \undefined \def \bcomment#1{#1} \fi
\ifx \oauthor \undefined \def \oauthor#1{#1} \fi
\ifx \citeauthoryear \undefined \def \citeauthoryear#1{#1} \fi
\ifx \endbibitem  \undefined \def \endbibitem {}\fi
\ifx \bconflocation  \undefined \def \bconflocation#1{#1} \fi
\ifx \arxivurl  \undefined \def \arxivurl#1{\textsf{#1}} \fi
\csname PreBibitemsHook\endcsname

\bibitem[\protect\citeauthoryear{{Abolmasov} and {Poutanen}}{2021}]{2021A&A...647A..45A}
\begin{barticle}
\bauthor{\bsnm{{Abolmasov}}, \binits{P.}},
\bauthor{\bsnm{{Poutanen}}, \binits{J.}}:
\bjtitle{Astronomy and Astrophysics}
\bvolume{647},
\bfpage{45}
(\byear{2021})
\end{barticle}
\endbibitem

\bibitem[\protect\citeauthoryear{{Abolmasov} et~al.}{2020}]{2020A&A...638A.142A}
\begin{barticle}
\bauthor{\bsnm{{Abolmasov}}, \binits{P.}},
\bauthor{\bsnm{{N{\"a}ttil{\"a}}}, \binits{J.}},
\bauthor{\bsnm{{Poutanen}}, \binits{J.}}:
\bjtitle{Astronomy and Astrophysics}
\bvolume{638},
\bfpage{142}
(\byear{2020})
\end{barticle}
\endbibitem

\bibitem[\protect\citeauthoryear{{Alpar} et~al.}{1992}]{1992A&A...257..627A}
\begin{barticle}
\bauthor{\bsnm{{Alpar}}, \binits{M.A.}},
\bauthor{\bsnm{{Hasinger}}, \binits{G.}},
\bauthor{\bsnm{{Shaham}}, \binits{J.}},
\bauthor{\bsnm{{Yancopoulos}}, \binits{S.}}:
\bjtitle{\aap}
\bvolume{257}(\bissue{2}),
\bfpage{627}
(\byear{1992})
\end{barticle}
\endbibitem

\bibitem[\protect\citeauthoryear{{Andrew} and {Purton}}{1968}]{1968Natur.218..855A}
\begin{barticle}
\bauthor{\bsnm{{Andrew}}, \binits{B.H.}},
\bauthor{\bsnm{{Purton}}, \binits{C.R.}}:
\bjtitle{\nat}
\bvolume{218}(\bissue{5144}),
\bfpage{855}
(\byear{1968})
\end{barticle}
\endbibitem

\bibitem[\protect\citeauthoryear{{Arnaud}}{1996}]{1996ASPC..101...17A}
\begin{bchapter}
\bauthor{\bsnm{{Arnaud}}, \binits{K.A.}}:
In: \beditor{\bsnm{{Jacoby}}, \binits{G.H.}},
\beditor{\bsnm{{Barnes}}, \binits{J.}} (eds.)
\bbtitle{Astronomical Data Analysis Software and Systems V}.
\bsertitle{Astronomical Society of the Pacific Conference Series},
vol. \bseriesno{101},
p. \bfpage{17}
(\byear{1996})
\end{bchapter}
\endbibitem

\bibitem[\protect\citeauthoryear{{Bałucińska-Church, M.} et~al.}{2010}]{refId0}
\begin{barticle}
\bauthor{\bsnm{{Bałucińska-Church, M.}}},
\bauthor{\bsnm{{Gibiec, A.}}},
\bauthor{\bsnm{{Jackson, N. K.}}},
\bauthor{\bsnm{{Church, M. J.}}}:
\bjtitle{A\&A}
\bvolume{512},
\bfpage{9}
(\byear{2010})
\end{barticle}
\endbibitem

\bibitem[\protect\citeauthoryear{{Bradshaw} et~al.}{1999}]{1999ApJ...512L.121B}
\begin{barticle}
\bauthor{\bsnm{{Bradshaw}}, \binits{C.F.}},
\bauthor{\bsnm{{Fomalont}}, \binits{E.B.}},
\bauthor{\bsnm{{Geldzahler}}, \binits{B.J.}}:
\bjtitle{\apjl}
\bvolume{512}(\bissue{2}),
\bfpage{121}
(\byear{1999})
\end{barticle}
\endbibitem

\bibitem[\protect\citeauthoryear{{Cackett} et~al.}{2010}]{2010ApJ...720..205C}
\begin{barticle}
\bauthor{\bsnm{{Cackett}}, \binits{E.M.}},
\bauthor{\bsnm{{Miller}}, \binits{J.M.}},
\bauthor{\bsnm{{Ballantyne}}, \binits{D.R.}},
\bauthor{\bsnm{{Barret}}, \binits{D.}},
\bauthor{\bsnm{{Bhattacharyya}}, \binits{S.}},
\bauthor{\bsnm{{Boutelier}}, \binits{M.}},
\bauthor{\bsnm{{Miller}}, \binits{M.C.}},
\bauthor{\bsnm{{Strohmayer}}, \binits{T.E.}},
\bauthor{\bsnm{{Wijnands}}, \binits{R.}}:
\bjtitle{\apj}
\bvolume{720}(\bissue{1}),
\bfpage{205}
(\byear{2010}).
\arxivurl{0908.1098}
\end{barticle}
\endbibitem

\bibitem[\protect\citeauthoryear{Chiranjeevi and Sriram}{2022}]{10.1093/mnras/stac2319}
\begin{barticle}
\bauthor{\bsnm{Chiranjeevi}, \binits{P.}},
\bauthor{\bsnm{Sriram}, \binits{K.}}:
\bjtitle{Monthly Notices of the Royal Astronomical Society}
\bvolume{516}(\bissue{2}),
\bfpage{2500}
(\byear{2022})
\end{barticle}
\endbibitem

\bibitem[\protect\citeauthoryear{{Chiranjeevi} et~al.}{2023}]{2023Ap&SS.368...77C}
\begin{barticle}
\bauthor{\bsnm{{Chiranjeevi}}, \binits{P.}},
\bauthor{\bsnm{{Sriram}}, \binits{K.}},
\bauthor{\bsnm{{Malu}}, \binits{S.}},
\bauthor{\bsnm{{Agrawal}}, \binits{V.K.}}:
\bjtitle{\apss}
\bvolume{368}(\bissue{9}),
\bfpage{77}
(\byear{2023})
\end{barticle}
\endbibitem

\bibitem[\protect\citeauthoryear{Church and Bałucińska-Church}{2004}]{10.1111/j.1365-2966.2004.07162.x}
\begin{barticle}
\bauthor{\bsnm{Church}, \binits{M.J.}},
\bauthor{\bsnm{Bałucińska-Church}, \binits{M.}}:
\bjtitle{Monthly Notices of the Royal Astronomical Society}
\bvolume{348}(\bissue{3}),
\bfpage{955}
(\byear{2004})
\end{barticle}
\endbibitem

\bibitem[\protect\citeauthoryear{Church et~al.}{2014}]{10.1093/mnras/stt2364}
\begin{barticle}
\bauthor{\bsnm{Church}, \binits{M.J.}},
\bauthor{\bsnm{Gibiec}, \binits{A.}},
\bauthor{\bsnm{Bałucińska-Church}, \binits{M.}}:
\bjtitle{Monthly Notices of the Royal Astronomical Society}
\bvolume{438}(\bissue{4}),
\bfpage{2784}
(\byear{2014})
\end{barticle}
\endbibitem

\bibitem[\protect\citeauthoryear{Church et~al.}{2012}]{Church:2012nb}
\begin{barticle}
\bauthor{\bsnm{Church}, \binits{M.J.}},
\bauthor{\bsnm{Gibiec}, \binits{A.}},
\bauthor{\bsnm{Balucinska-Church}, \binits{M.}},
\bauthor{\bsnm{Jackson}, \binits{N.K.}}:
\bjtitle{Astron. Astrophys.}
\bvolume{546},
\bfpage{35}
(\byear{2012}).
\arxivurl{1208.4723}
\end{barticle}
\endbibitem

\bibitem[\protect\citeauthoryear{Ding et~al.}{2023}]{Ding_2023}
\begin{barticle}
\bauthor{\bsnm{Ding}, \binits{G.Q.}},
\bauthor{\bsnm{Qu}, \binits{J.L.}},
\bauthor{\bsnm{Song}, \binits{L.M.}},
\bauthor{\bsnm{Huang}, \binits{Y.}},
\bauthor{\bsnm{Zhang}, \binits{S.}},
\bauthor{\bsnm{Bu}, \binits{Q.C.}},
\bauthor{\bsnm{Ge}, \binits{M.Y.}},
\bauthor{\bsnm{Li}, \binits{X.B.}},
\bauthor{\bsnm{Tao}, \binits{L.}},
\bauthor{\bsnm{Ma}, \binits{X.}},
\bauthor{\bsnm{Chen}, \binits{Y.P.}},
\bauthor{\bsnm{Zhang}, \binits{L.}},
\bauthor{\bsnm{Yan}, \binits{W.M.}},
\bauthor{\bsnm{Tuo}, \binits{Y.L.}},
\bauthor{\bsnm{Fu}, \binits{Y.C.}},
\bauthor{\bsnm{Xiao}, \binits{S.H.}},
\bauthor{\bsnm{Yang}, \binits{Z.X.}},
\bauthor{\bsnm{Liu}, \binits{H.X.}}:
\bjtitle{The Astrophysical Journal}
\bvolume{950}(\bissue{1}),
\bfpage{69}
(\byear{2023})
\end{barticle}
\endbibitem

\bibitem[\protect\citeauthoryear{{Fomalont} et~al.}{2001}]{2001ApJ...553L..27F}
\begin{barticle}
\bauthor{\bsnm{{Fomalont}}, \binits{E.B.}},
\bauthor{\bsnm{{Geldzahler}}, \binits{B.J.}},
\bauthor{\bsnm{{Bradshaw}}, \binits{C.F.}}:
\bjtitle{\apjl}
\bvolume{553}(\bissue{1}),
\bfpage{27}
(\byear{2001}).
\arxivurl{astro-ph/0104325}
\end{barticle}
\endbibitem

\bibitem[\protect\citeauthoryear{Fomalont et~al.}{2001}]{Fomalont_2001}
\begin{barticle}
\bauthor{\bsnm{Fomalont}, \binits{E.B.}},
\bauthor{\bsnm{Geldzahler}, \binits{B.J.}},
\bauthor{\bsnm{Bradshaw}, \binits{C.F.}}:
\bjtitle{The Astrophysical Journal}
\bvolume{558}(\bissue{1}),
\bfpage{283}
(\byear{2001})
\end{barticle}
\endbibitem

\bibitem[\protect\citeauthoryear{Fortner et~al.}{1989}]{Fortner1989}
\begin{barticle}
\bauthor{\bsnm{Fortner}, \binits{B.}},
\bauthor{\bsnm{Lamb}, \binits{F.K.}},
\bauthor{\bsnm{Miller}, \binits{G.S.}}:
\bjtitle{Nature}
\bvolume{342}(\bissue{6251}),
\bfpage{775}
(\byear{1989})
\end{barticle}
\endbibitem

\bibitem[\protect\citeauthoryear{Giacconi et~al.}{1962}]{PhysRevLett.9.439}
\begin{barticle}
\bauthor{\bsnm{Giacconi}, \binits{R.}},
\bauthor{\bsnm{Gursky}, \binits{H.}},
\bauthor{\bsnm{Paolini}, \binits{F.R.}},
\bauthor{\bsnm{Rossi}, \binits{B.B.}}:
\bjtitle{Phys. Rev. Lett.}
\bvolume{9},
\bfpage{439}
(\byear{1962})
\end{barticle}
\endbibitem

\bibitem[\protect\citeauthoryear{Harikrishna and Sriram}{2022}]{10.1093/mnras/stac2527}
\begin{barticle}
\bauthor{\bsnm{Harikrishna}, \binits{S.}},
\bauthor{\bsnm{Sriram}, \binits{K.}}:
\bjtitle{Monthly Notices of the Royal Astronomical Society}
\bvolume{516}(\bissue{4}),
\bfpage{5148}
(\byear{2022})
\end{barticle}
\endbibitem

\bibitem[\protect\citeauthoryear{{Hasinger} and {van der Klis}}{1989}]{1989A&A...225...79H}
\begin{barticle}
\bauthor{\bsnm{{Hasinger}}, \binits{G.}},
\bauthor{\bsnm{{van der Klis}}, \binits{M.}}:
\bjtitle{\aap}
\bvolume{225},
\bfpage{79}
(\byear{1989})
\end{barticle}
\endbibitem

\bibitem[\protect\citeauthoryear{{Hasinger} et~al.}{1990}]{1990A&A...235..131H}
\begin{barticle}
\bauthor{\bsnm{{Hasinger}}, \binits{G.}},
\bauthor{\bsnm{{van der Klis}}, \binits{M.}},
\bauthor{\bsnm{{Ebisawa}}, \binits{K.}},
\bauthor{\bsnm{{Dotani}}, \binits{T.}},
\bauthor{\bsnm{{Mitsuda}}, \binits{K.}}:
\bjtitle{\aap}
\bvolume{235},
\bfpage{131}
(\byear{1990})
\end{barticle}
\endbibitem

\bibitem[\protect\citeauthoryear{Jahoda et~al.}{2006}]{Jahoda_2006}
\begin{barticle}
\bauthor{\bsnm{Jahoda}, \binits{K.}},
\bauthor{\bsnm{Markwardt}, \binits{C.B.}},
\bauthor{\bsnm{Radeva}, \binits{Y.}},
\bauthor{\bsnm{Rots}, \binits{A.H.}},
\bauthor{\bsnm{Stark}, \binits{M.J.}},
\bauthor{\bsnm{Swank}, \binits{J.H.}},
\bauthor{\bsnm{Strohmayer}, \binits{T.E.}},
\bauthor{\bsnm{Zhang}, \binits{W.}}:
\bjtitle{The Astrophysical Journal Supplement Series}
\bvolume{163}(\bissue{2}),
\bfpage{401}
(\byear{2006})
\end{barticle}
\endbibitem

\bibitem[\protect\citeauthoryear{Jia et~al.}{2023}]{jia2023study}
\begin{barticle}
\bauthor{\bsnm{Jia}, \binits{S.}},
\bauthor{\bsnm{Qu}, \binits{J.}},
\bauthor{\bsnm{Lu}, \binits{F.}},
\bauthor{\bsnm{Zhang}, \binits{L.}},
\bauthor{\bsnm{Zhang}, \binits{S.}},
\bauthor{\bsnm{Song}, \binits{L.}},
\bauthor{\bsnm{Zhang}, \binits{S.}},
\bauthor{\bsnm{Huang}, \binits{Y.}},
\bauthor{\bsnm{Ma}, \binits{X.}},
\bauthor{\bsnm{Tao}, \binits{L.}}, \betal:
\bjtitle{Monthly Notices of the Royal Astronomical Society}
\bvolume{521}(\bissue{3}),
\bfpage{4792}
(\byear{2023})
\end{barticle}
\endbibitem

\bibitem[\protect\citeauthoryear{Kuulkers et~al.}{1997}]{kuulkers1997gx}
\begin{barticle}
\bauthor{\bsnm{Kuulkers}, \binits{E.}},
\bauthor{\bsnm{Kuulkers}, \binits{E.}},
\bauthor{\bparticle{van~der} \bsnm{Klis}, \binits{M.}},
\bauthor{\bsnm{Oosterbroek}, \binits{T.}},
\bauthor{\bparticle{van} \bsnm{Paradijs}, \binits{J.}},
\bauthor{\bsnm{Lewin}, \binits{W.}}:
\bjtitle{Monthly Notices of the Royal Astronomical Society}
\bvolume{287}(\bissue{3}),
\bfpage{495}
(\byear{1997})
\end{barticle}
\endbibitem

\bibitem[\protect\citeauthoryear{Lamb}{1989}]{lamb198923rd}
\begin{bchapter}
\bauthor{\bsnm{Lamb}, \binits{F.}}:
\bctitle{23rd eslab symp., two topics in x-ray astronomy}.
(\byear{1989}).
\bcomment{ESA}
\end{bchapter}
\endbibitem

\bibitem[\protect\citeauthoryear{{Lei} et~al.}{2008}]{2008ApJ...677..461L}
\begin{barticle}
\bauthor{\bsnm{{Lei}}, \binits{Y.J.}},
\bauthor{\bsnm{{Qu}}, \binits{J.L.}},
\bauthor{\bsnm{{Song}}, \binits{L.M.}},
\bauthor{\bsnm{{Zhang}}, \binits{C.M.}},
\bauthor{\bsnm{{Zhang}}, \binits{S.}},
\bauthor{\bsnm{{Zhang}}, \binits{F.}},
\bauthor{\bsnm{{Wang}}, \binits{J.M.}},
\bauthor{\bsnm{{Li}}, \binits{Z.B.}},
\bauthor{\bsnm{{Zhang}}, \binits{G.B.}}:
\bjtitle{The Astrophysical Journal}
\bvolume{677}(\bissue{1}),
\bfpage{461}
(\byear{2008})
\end{barticle}
\endbibitem

\bibitem[\protect\citeauthoryear{Malu et~al.}{2020}]{malu2020coronal}
\begin{barticle}
\bauthor{\bsnm{Malu}, \binits{S.}},
\bauthor{\bsnm{Sriram}, \binits{K.}},
\bauthor{\bsnm{Agrawal}, \binits{V.}}:
\bjtitle{Monthly Notices of the Royal Astronomical Society}
\bvolume{499}(\bissue{2}),
\bfpage{2214}
(\byear{2020})
\end{barticle}
\endbibitem

\bibitem[\protect\citeauthoryear{Malu et~al.}{2021}]{malu2021investigating}
\begin{barticle}
\bauthor{\bsnm{Malu}, \binits{S.}},
\bauthor{\bsnm{Harikrishna}, \binits{S.}},
\bauthor{\bsnm{Sriram}, \binits{K.}},
\bauthor{\bsnm{Agrawal}, \binits{V.K.}}:
\bjtitle{Astrophysics and Space Science}
\bvolume{366}(\bissue{9}),
\bfpage{1}
(\byear{2021})
\end{barticle}
\endbibitem

\bibitem[\protect\citeauthoryear{Marino et~al.}{2023}]{marino2023accretion}
\begin{barticle}
\bauthor{\bsnm{Marino}, \binits{A.}},
\bauthor{\bsnm{Russell}, \binits{T.}},
\bauthor{\bsnm{Del~Santo}, \binits{M.}},
\bauthor{\bsnm{Beri}, \binits{A.}},
\bauthor{\bsnm{Sanna}, \binits{A.}},
\bauthor{\bsnm{Coti~Zelati}, \binits{F.}},
\bauthor{\bsnm{Degenaar}, \binits{N.}},
\bauthor{\bsnm{Altamirano}, \binits{D.}},
\bauthor{\bsnm{Ambrosi}, \binits{E.}},
\bauthor{\bsnm{Anitra}, \binits{A.}}, \betal:
\bjtitle{Monthly Notices of the Royal Astronomical Society}
\bvolume{525}(\bissue{2}),
\bfpage{2366}
(\byear{2023})
\end{barticle}
\endbibitem

\bibitem[\protect\citeauthoryear{M{\'e}ndez et~al.}{2022}]{mendez2022coupling}
\begin{barticle}
\bauthor{\bsnm{M{\'e}ndez}, \binits{M.}},
\bauthor{\bsnm{Karpouzas}, \binits{K.}},
\bauthor{\bsnm{Garc{\'\i}a}, \binits{F.}},
\bauthor{\bsnm{Zhang}, \binits{L.}},
\bauthor{\bsnm{Zhang}, \binits{Y.}},
\bauthor{\bsnm{Belloni}, \binits{T.M.}},
\bauthor{\bsnm{Altamirano}, \binits{D.}}:
\bjtitle{Nature Astronomy}
\bvolume{6}(\bissue{5}),
\bfpage{577}
(\byear{2022})
\end{barticle}
\endbibitem

\bibitem[\protect\citeauthoryear{Migliari et~al.}{2007}]{migliari2007linking}
\begin{barticle}
\bauthor{\bsnm{Migliari}, \binits{S.}},
\bauthor{\bsnm{Miller-Jones}, \binits{J.}},
\bauthor{\bsnm{Fender}, \binits{R.}},
\bauthor{\bsnm{Homan}, \binits{J.}},
\bauthor{\bsnm{Di~Salvo}, \binits{T.}},
\bauthor{\bsnm{Rothschild}, \binits{R.}},
\bauthor{\bsnm{Rupen}, \binits{M.}},
\bauthor{\bsnm{Tomsick}, \binits{J.}},
\bauthor{\bsnm{Wijnands}, \binits{R.}},
\bauthor{\bparticle{Van~der} \bsnm{Klis}, \binits{M.}}:
\bjtitle{The Astrophysical Journal}
\bvolume{671}(\bissue{1}),
\bfpage{706}
(\byear{2007})
\end{barticle}
\endbibitem

\bibitem[\protect\citeauthoryear{Miller and Lamb}{1992}]{miller1992energy}
\begin{barticle}
\bauthor{\bsnm{Miller}, \binits{G.S.}},
\bauthor{\bsnm{Lamb}, \binits{F.K.}}:
\bjtitle{Astrophysical Journal, Part 1 (ISSN 0004-637X), vol. 388, April 1, 1992, p. 541-554. Research supported by DOE.}
\bvolume{388},
\bfpage{541}
(\byear{1992})
\end{barticle}
\endbibitem

\bibitem[\protect\citeauthoryear{Motta and Fender}{2019}]{motta2019connection}
\begin{barticle}
\bauthor{\bsnm{Motta}, \binits{S.}},
\bauthor{\bsnm{Fender}, \binits{R.}}:
\bjtitle{Monthly Notices of the Royal Astronomical Society}
\bvolume{483}(\bissue{3}),
\bfpage{3686}
(\byear{2019})
\end{barticle}
\endbibitem

\bibitem[\protect\citeauthoryear{Ng et~al.}{2024}]{ng2024x}
\begin{barticle}
\bauthor{\bsnm{Ng}, \binits{M.}},
\bauthor{\bsnm{Hughes}, \binits{A.K.}},
\bauthor{\bsnm{Homan}, \binits{J.}},
\bauthor{\bsnm{Miller}, \binits{J.M.}},
\bauthor{\bsnm{Pike}, \binits{S.N.}},
\bauthor{\bsnm{Altamirano}, \binits{D.}},
\bauthor{\bsnm{Bult}, \binits{P.}},
\bauthor{\bsnm{Chakrabarty}, \binits{D.}},
\bauthor{\bsnm{Buisson}, \binits{D.}},
\bauthor{\bsnm{Coughenour}, \binits{B.M.}}, \betal:
\bjtitle{The Astrophysical Journal}
\bvolume{966}(\bissue{2}),
\bfpage{232}
(\byear{2024})
\end{barticle}
\endbibitem

\bibitem[\protect\citeauthoryear{Popham and Sunyaev}{2001}]{popham2001accretion}
\begin{barticle}
\bauthor{\bsnm{Popham}, \binits{R.}},
\bauthor{\bsnm{Sunyaev}, \binits{R.}}:
\bjtitle{The Astrophysical Journal}
\bvolume{547}(\bissue{1}),
\bfpage{355}
(\byear{2001})
\end{barticle}
\endbibitem

\bibitem[\protect\citeauthoryear{{Revnivtsev} and {Gilfanov}}{2006}]{2006A&A...453..253R}
\begin{barticle}
\bauthor{\bsnm{{Revnivtsev}}, \binits{M.G.}},
\bauthor{\bsnm{{Gilfanov}}, \binits{M.R.}}:
\bjtitle{Astronomy and Astrophysics}
\bvolume{453}(\bissue{1}),
\bfpage{253}
(\byear{2006})
\end{barticle}
\endbibitem

\bibitem[\protect\citeauthoryear{Russell et~al.}{2016}]{russell2016reproducible}
\begin{barticle}
\bauthor{\bsnm{Russell}, \binits{T.D.}},
\bauthor{\bsnm{Miller-Jones}, \binits{J.C.}},
\bauthor{\bsnm{Sivakoff}, \binits{G.R.}},
\bauthor{\bsnm{Altamirano}, \binits{D.}},
\bauthor{\bsnm{O'Brien}, \binits{T.J.}},
\bauthor{\bsnm{Page}, \binits{K.L.}},
\bauthor{\bsnm{Templeton}, \binits{M.R.}},
\bauthor{\bsnm{K{\"o}rding}, \binits{E.}},
\bauthor{\bsnm{Knigge}, \binits{C.}},
\bauthor{\bsnm{Rupen}, \binits{M.}}, \betal:
\bjtitle{Monthly Notices of the Royal Astronomical Society}
\bvolume{460}(\bissue{4}),
\bfpage{3720}
(\byear{2016})
\end{barticle}
\endbibitem

\bibitem[\protect\citeauthoryear{Schulz et~al.}{2009}]{schulz2009heating}
\begin{barticle}
\bauthor{\bsnm{Schulz}, \binits{N.}},
\bauthor{\bsnm{Huenemoerder}, \binits{D.}},
\bauthor{\bsnm{Ji}, \binits{L.}},
\bauthor{\bsnm{Nowak}, \binits{M.}},
\bauthor{\bsnm{Yao}, \binits{Y.}},
\bauthor{\bsnm{Canizares}, \binits{C.}}:
\bjtitle{The Astrophysical Journal}
\bvolume{692}(\bissue{2}),
\bfpage{80}
(\byear{2009})
\end{barticle}
\endbibitem

\bibitem[\protect\citeauthoryear{Sriram et~al.}{2012}]{sriram2012anti}
\begin{barticle}
\bauthor{\bsnm{Sriram}, \binits{K.}},
\bauthor{\bsnm{Choi}, \binits{C.}},
\bauthor{\bsnm{Rao}, \binits{A.}}:
\bjtitle{The Astrophysical Journal Supplement Series}
\bvolume{200}(\bissue{2}),
\bfpage{16}
(\byear{2012})
\end{barticle}
\endbibitem

\bibitem[\protect\citeauthoryear{Sriram et~al.}{2019}]{sriram2019constraining}
\begin{barticle}
\bauthor{\bsnm{Sriram}, \binits{K.}},
\bauthor{\bsnm{Malu}, \binits{S.}},
\bauthor{\bsnm{Choi}, \binits{C.}}:
\bjtitle{The Astrophysical Journal Supplement Series}
\bvolume{244}(\bissue{1}),
\bfpage{5}
(\byear{2019})
\end{barticle}
\endbibitem

\bibitem[\protect\citeauthoryear{Sriram et~al.}{2011}]{sriram2011coupled}
\begin{barticle}
\bauthor{\bsnm{Sriram}, \binits{K.}},
\bauthor{\bsnm{Rao}, \binits{A.}},
\bauthor{\bsnm{Choi}, \binits{C.}}:
\bjtitle{The Astrophysical Journal Letters}
\bvolume{743}(\bissue{2}),
\bfpage{31}
(\byear{2011})
\end{barticle}
\endbibitem

\bibitem[\protect\citeauthoryear{Sriram et~al.}{2021}]{sriram2021understanding}
\begin{barticle}
\bauthor{\bsnm{Sriram}, \binits{K.}},
\bauthor{\bsnm{Chiranjeevi}, \binits{P.}},
\bauthor{\bsnm{Malu}, \binits{S.}},
\bauthor{\bsnm{Agrawal}, \binits{V.}}:
\bjtitle{Journal of Astrophysics and Astronomy}
\bvolume{42}(\bissue{2}),
\bfpage{1}
(\byear{2021})
\end{barticle}
\endbibitem

\bibitem[\protect\citeauthoryear{Steeghs and Casares}{2002}]{steeghs2002mass}
\begin{barticle}
\bauthor{\bsnm{Steeghs}, \binits{D.}},
\bauthor{\bsnm{Casares}, \binits{J.}}:
\bjtitle{The Astrophysical Journal}
\bvolume{568}(\bissue{1}),
\bfpage{273}
(\byear{2002})
\end{barticle}
\endbibitem

\bibitem[\protect\citeauthoryear{Sudha et~al.}{2024}]{sudha2024spectro}
\begin{barticle}
\bauthor{\bsnm{Sudha}, \binits{M.}},
\bauthor{\bsnm{Ludlam}, \binits{R.M.}},
\bauthor{\bsnm{Altamirano}, \binits{D.}},
\bauthor{\bsnm{Cackett}, \binits{E.M.}},
\bauthor{\bsnm{Hare}, \binits{J.}}:
\bjtitle{The Astrophysical Journal}
\bvolume{978}(\bissue{1}),
\bfpage{75}
(\byear{2024})
\end{barticle}
\endbibitem

\bibitem[\protect\citeauthoryear{{Timmer} and {K{\"o}nig}}{1995}]{1995A&A...300..707T}
\begin{barticle}
\bauthor{\bsnm{{Timmer}}, \binits{J.}},
\bauthor{\bsnm{{K{\"o}nig}}, \binits{M.}}:
\bjtitle{\aap}
\bvolume{300},
\bfpage{707}
(\byear{1995})
\end{barticle}
\endbibitem

\bibitem[\protect\citeauthoryear{Titarchuk et~al.}{2014}]{titarchuk2014x}
\begin{barticle}
\bauthor{\bsnm{Titarchuk}, \binits{L.}},
\bauthor{\bsnm{Seifina}, \binits{E.}},
\bauthor{\bsnm{Shrader}, \binits{C.}}:
\bjtitle{The Astrophysical Journal}
\bvolume{789}(\bissue{2}),
\bfpage{98}
(\byear{2014})
\end{barticle}
\endbibitem

\bibitem[\protect\citeauthoryear{Titarchuk et~al.}{2001}]{titarchuk2001normal}
\begin{barticle}
\bauthor{\bsnm{Titarchuk}, \binits{L.}},
\bauthor{\bsnm{Bradshaw}, \binits{C.}},
\bauthor{\bsnm{Geldzahler}, \binits{B.}},
\bauthor{\bsnm{Fomalont}, \binits{E.}}:
\bjtitle{The Astrophysical Journal}
\bvolume{555}(\bissue{1}),
\bfpage{45}
(\byear{2001})
\end{barticle}
\endbibitem

\bibitem[\protect\citeauthoryear{Van~der Klis}{2006}]{van2006compact}
\begin{bbook}
\bauthor{\bparticle{Van~der} \bsnm{Klis}, \binits{M.}}:
\bbtitle{Compact Stellar X-ray Sources}.
\bpublisher{Cambridge University Press}, \blocation{???}
(\byear{2006})
\end{bbook}
\endbibitem

\bibitem[\protect\citeauthoryear{{Wijnands} and {van der Klis}}{1999}]{1999ApJ...514..939W}
\begin{barticle}
\bauthor{\bsnm{{Wijnands}}, \binits{R.}},
\bauthor{\bsnm{{van der Klis}}, \binits{M.}}:
\bjtitle{The Astrophysical Journal}
\bvolume{514}(\bissue{2}),
\bfpage{939}
(\byear{1999})
\end{barticle}
\endbibitem

\bibitem[\protect\citeauthoryear{Yin et~al.}{2007}]{yin2007correlations}
\begin{barticle}
\bauthor{\bsnm{Yin}, \binits{H.}},
\bauthor{\bsnm{Zhang}, \binits{C.}},
\bauthor{\bsnm{Zhao}, \binits{Y.}},
\bauthor{\bsnm{Lei}, \binits{Y.}},
\bauthor{\bsnm{Qu}, \binits{J.}},
\bauthor{\bsnm{Song}, \binits{L.}},
\bauthor{\bsnm{Zhang}, \binits{F.}}:
\bjtitle{Astronomy \& Astrophysics}
\bvolume{471}(\bissue{2}),
\bfpage{381}
(\byear{2007})
\end{barticle}
\endbibitem

\bibitem[\protect\citeauthoryear{Zdziarski et~al.}{1996}]{zdziarski1996broad}
\begin{barticle}
\bauthor{\bsnm{Zdziarski}, \binits{A.A.}},
\bauthor{\bsnm{Johnson}, \binits{W.N.}},
\bauthor{\bsnm{Magdziarz}, \binits{P.}}:
\bjtitle{Monthly Notices of the Royal Astronomical Society}
\bvolume{283}(\bissue{1}),
\bfpage{193}
(\byear{1996})
\end{barticle}
\endbibitem

\end{thebibliography}

\end{document}